\documentclass[twocolumn,showpacs,preprintnumbers,amsmath,amssymb]{revtex4}
\usepackage{amsmath,amssymb,graphics,epsfig,subfigure}

\begin{document}
\newcommand {\nn}    {\nonumber}

\title{Localization and Mass Spectra of Fermions on Symmetric and Asymmetric Thick Branes}

\author{Yu-Xiao Liu\footnote{liuyx@lzu.edu.cn}\footnote{Corresponding author.},
        Chun-E Fu\footnote{fuche08@lzu.cn},
        Li Zhao\footnote{lizhao@lzu.edu.cn},
        Yi-Shi Duan\footnote{ysduan@lzu.edu.cn}}
\affiliation{Institute of Theoretical Physics,
      Lanzhou University, Lanzhou 730000, People's Republic of China}

\begin{abstract}
A three-parameter (positive odd integer $s$, thickness factor
$\lambda$, and asymmetry factor $a$) family of asymmetric thick
brane solutions in five dimensions were constructed from a
two-parameter ($s$ and $\lambda$) family of symmetric ones in [R.
Guerrero, R.O. Rodriguez, and R. Torrealba, Phys. Rev. D
\textbf{72}, 124012 (2005).]. The values $s=1$ and $s\geq3$
correspond to single branes and double branes, respectively. These
branes have very rich inner structure. In this paper, by
presenting the mass-independent potentials of Kaluza--Klein (KK)
modes in the corresponding Schr\"{o}dinger equations, we
investigate the localization and mass spectra of fermions on the
symmetric and asymmetric thick branes in an AdS background. In
order to analyze the effect of gravity-fermion interaction (i.e.,
the effect of the inner structure of the branes) and
scalar-fermion interaction to the spectrum of fermion KK modes, we
consider three kinds of typical kink-fermion couplings. The
spectra of left chiral fermions for these couplings are consisted
of a bound zero mode and a series of gapless continuous massive KK
modes, some discrete bound KK modes including zero mode (exist
mass gaps) and a series of continuous massive KK modes, infinite
discrete bound KK modes, respectively. The structures of the
spectra are investigated in detail.
\end{abstract}




\maketitle

\section{Introduction}

The suggestion that our observed four-dimensional world is a brane
embedded in a higher-dimensional space-time
\cite{Rubakov1983,Akama1983,ADD,rs,Lykken} can provide new
insights for solving gauge hierarchy problem and cosmological
constant problem, etc. In the framework of brane scenarios,
gravity is free to propagate in all dimensions, while all the
matter fields are confined to a 3--brane. By introducing large
extra dimensions, Arkani-Hamed-Dimopoulos-Dvali (ADD) brane model
\cite{ADD} drops the fundamental Planck scale to Tev. However, it
introduces intermediate mass scales corresponding to the large
extra dimensions between Planck and Tev scales. In Ref. \cite{rs},
an alternative scenario, Randall-Sundrum (RS) warped brane model,
had been proposed. In this scenario, the internal manifold does
not need to be compactified to the Planck scale anymore and the
exponential warp factor in the metric can generate a large
hierarchy of scales, which are reasons why this new brane model
has attracted so much attention.

Generalizations and extensions of RS brane model have been proposed
for examples in Ref.
\cite{ExtensionRS,Starkman2001,SenGuptaPRL2007}. In Ref.
\cite{Starkman2001}, the model with extra dimensions composed of a
compact hyperbolic manifold is free of usual problems that plague
the original ADD model and shares many common features with RS
model. Recently, RS model is generalized to higher dimensions for a
multiply space-time with negative cosmological constant
\cite{SenGuptaPRL2007}. In this generalized scenario the observed
hierarchy in the masses of standard model fermions can be explained
geometrically without invoking any further hierarchy among the
various moduli provided the warping is large in one direction and
small in the other.

In RS warped scenario, however, the modulus, namely the brane
separation, is not stable. Goldberger and Wise showed that it can
be stabilized by introducing a scalar field in the bulk \cite{GW}.
A bulk scalar also provides us with a way of generating the brane
as a domain wall (thick brane) in five dimensions. Considering our
four-dimensional Universe as an infinitely thin domain wall is an
idealization. It is for this reason that an increasing interest
has been focused on the study of thick brane scenarios based on
gravity coupled to scalars in higher dimensional space-time
\cite{dewolfe,GremmPLB2000,gremm,Csaki,CamposPRL2002,varios,Guerrero2002,ThickBraneDzhunushaliev,ThickBraneBazeia,ShtanovJCAP2009}.
A virtue of these models is that the branes can be obtained
naturally rather than introduced by hand. In most thick brane
scenarios, the scalar field is a standard topological kink
interpolating between the minima of a potential with spontaneously
broken symmetry. For a comprehensive review on thick brane
solutions and related topics please see Ref.
\cite{ThickBraneReview}.

In brane world scenarios, an important problem is localization of
various bulk fields on a brane by a natural mechanism. Especially,
the localization of spin half fermions on thick branes is very
interesting. Localizing fermions on branes or defects requires us
to introduce other interactions besides gravity. Recently,
localization mechanisms on a domain wall for fermions have been
extensively analyzed in Ref. \cite{Volkas2007}. There are some
other backgrounds, for example, gauge field
\cite{Parameswaran0608074,LiuJHEP2007}, supergravity \cite{Mario}
and vortex background
\cite{LiuNPB2007,LiuVortexFermion,Rafael200803,StojkovicPRD},
could be considered. Localization of fermions in general
space-times had been studied for example in
\cite{RandjbarPLB2000}. In five dimensions, with the
scalar--fermion coupling, there may exist a single bound state and
a continuous gapless spectrum of massive fermion Kaluza--Klein
(KK) states
\cite{ThickBrane1,ThickBrane2,ThickBrane3,Liu0708,20082009}, while
for some other brane models, there exist finite discrete KK states
(mass gaps) and a continuous gapless spectrum starting at a
positive $m^2$ \cite{ThickBrane4,Liu0803,0803.1458}. In Ref.
\cite{DubovskyPRD2000}, it was found that fermions can escape into
the bulk by tunnelling, and the rate depends on the parameters of
the scalar potential. In Ref. \cite{KoleyCQG2005}, the authors
obtained trapped discrete massive fermion states on the brane,
which in fact are quasibound and have a finite probability of
escaping into the bulk.

In Ref. \cite{LiuJCAP2009}, localization and mass spectra of
various bulk matter fields including fermions on symmetric and
asymmetric de Sitter thick single branes were investigated. It was
shown that the massless modes of scalars and vectors are separated
by a mass gap from the continuous modes. The asymmetry may
increase the number of the bound KK modes of scalars but does not
change that of vectors. The localization property of spin 1/2
fermions is dependent on the coupling of fermions and the
background scalar $\eta\bar{\Psi}F(\phi)\Psi$. For the usual
Yukawa coupling with $F(\phi(z))=\phi(z){\sim}\arctan(\sinh z)$ (a
usual kink which is almost a constant at large $z$), the fermion
zero mode can not be localized on the branes. For the
scalar-fermion coupling with $F(\phi(z))$ a kink likes $\sinh z$,
which increases quickly with $z$, there exist some discrete bound
KK modes and a series of continuous ones, and one of the zero
modes of left and right fermions is localized on the branes
strongly. The asymmetry reduces the number of the bound fermion KK
modes.

Fermions on symmetric and asymmetric double branes have been
reported in Ref. \cite{Guerrero2006}. These double branes are stable
Bogomol'nyi-Prasad-Sommerfeld (BPS) thick walls with two sub walls
located at their edges. It was shown that, for the symmetric brane,
the zero modes of fermions coupled to the scalar field through
Yukawa interactions and gravitons are not peaked at the center of
the brane, but instead a constant between the two sub branes.
However, in the asymmetric scenario, as a consequence of the
asymmetry, fermions are localized on one of the sub walls, while the
gravitons are localized on another sub wall. Hence a large hierarchy
between the Planck and the weak scales can be produced.

In Ref. \cite{asymdSBrane2}, a three-parameter family of asymmetric
thick brane solutions in five dimensions (including single branes
and double branes) were constructed from a two-parameter family of
symmetric ones given in Refs.
\cite{MelfoPRD2003,GregoryPRD2002,BazeiaJCAP2004}. These branes have
very rich inner structure. In this paper, we will investigate the
localization problem and the mass spectra of fermions on the
symmetric and asymmetric thick branes for three kinds of typical
kink-fermion couplings in detail. It will be shown that the
localization properties on asymmetric branes are very different from
these given in Refs. \cite{LiuJCAP2009} and \cite{Guerrero2006}. The
mass spectra of fermions are determined by the inner structures of
the branes and the scalar-fermion couplings. The paper is organized
as follows: In Sec. \ref{SecModel}, we first give a brief review of
the symmetric and asymmetric double thick branes in an AdS
background. Then, in Sec. \ref{SecLocalize}, we study localization
of spin half fermions on the thick branes with different types of
scalar-fermion interactions by presenting the shapes of the
potentials of the corresponding Schr\"{o}dinger problem. Finally, a
brief discussion and conclusion are presented in Sec.
\ref{secConclusion}.

\section{Review of the symmetric and asymmetric thick branes}
\label{SecModel}

Let us consider thick branes arising from a real scalar field $\phi$
with a scalar potential $V(\phi)$. The action for such a system is
given by
\begin{equation}
S = \int d^5 x \sqrt{-g}\left [ \frac{1}{2\kappa_5^2} R-\frac{1}{2}
g^{MN}\partial_M \phi
\partial_N \phi - V(\phi) \right ],
\label{action}
\end{equation}
where $R$ is the scalar curvature and $\kappa_5^2=8 \pi G_5$ with
$G_5$ the five-dimensional Newton constant. Here we set
$\kappa_5=1$. The line-element for a five-dimensional space-time
is assumed as
\begin{eqnarray}
 ds^2&=&\text{e}^{2A(z)}\big(\eta_{\mu\nu}dx^\mu dx^\nu
          + dz^2\big), \label{linee}
\end{eqnarray}
where $\text{e}^{2A(z)}$ is the warp factor and $z$ stands for the
extra coordinate. The scalar field is considered to be a function of
$z$ only, i.e., $\phi=\phi(z)$. In the model, the potential could
provide a thick brane realization, and the soliton configuration of
the scalar field dynamically generates the domain wall configuration
with warped geometry. The field equations generated from the action
(\ref{action}) with the ansatz (\ref{linee}) reduce to the following
coupled nonlinear differential equations
\begin{eqnarray}
\phi'^2 & = & 3(A'^2-A''), \\
V(\phi) & = & \frac{3}{2}  (-3A'^2-A'') e^{-2A},\\
\frac{dV(\phi)}{d\phi} &  = & (3A'\phi'+\phi'')e^{-2A},
\end{eqnarray}
where the prime denotes derivative with respect to $z$. Now we
consider static double thick branes in an AdS background. A
two-parameter family of symmetric double thick branes in five
dimensions for the potential
\begin{eqnarray}
 V(\phi)&=&\frac{3}{2} \lambda^2
  \sin^{2-\frac{2}{s}} (\phi/\phi_{0})
  \cos^2(\phi/\phi_{0}) \nonumber \\
  &&\times\left[2s-1-4\tan^2(\phi/\phi_{0})\right],
 \label{potencialdoble1}
\end{eqnarray}
were presented and discussed in Refs.
\cite{MelfoPRD2003,GregoryPRD2002}:
\begin{eqnarray}
 e^{2A}&=&\frac{1}{\left[1+(\lambda z)^{2s}\right]^{1/s}},
         \label{e2A3} \\
 \phi~~&=&\phi_{0}\arctan(\lambda z)^s,
         \label{phi3}
\end{eqnarray}
where $\phi_{0}={\sqrt{3(2s-1)}}/{s}$, $\lambda$ is a positive real
constant, and $s$ is a positive odd integer. This solution
represents a family of plane symmetric static single ($s=1$) or
double ($s>1$) domain wall space-times, being asymptotically AdS$_5$
with a cosmological constant $-6\lambda^2$. Similar solutions can be
found in \cite{BazeiaJCAP2004}.

\begin{figure*}[htb]
\begin{center}
\includegraphics[width=4.8cm,height=3cm]{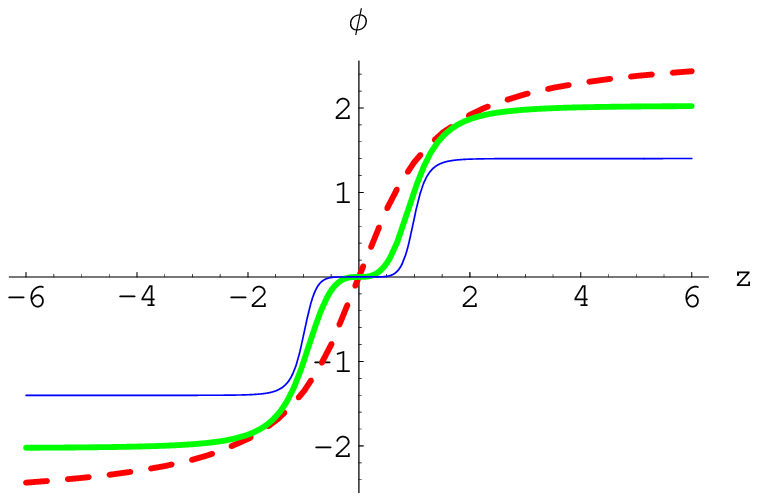}
\includegraphics[width=4.8cm,height=3cm]{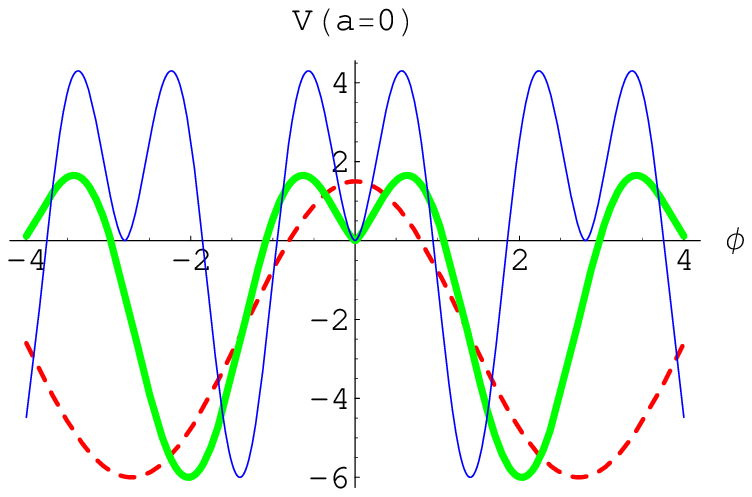}
\includegraphics[width=4.8cm,height=3cm]{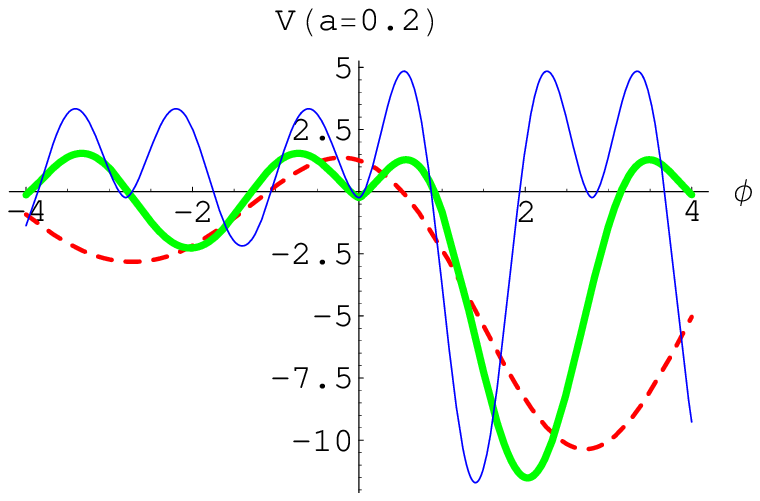}
\end{center}
\vskip -4mm \caption{The shapes of the kink $\phi(z)$ and the
potential $V(\phi)$ for $a=0$ and 0.2. The parameters are set to
$\lambda=1$, $s=1$ for red dashed lines, $s=3$ for green thick
lines, and $s=7$ for blue thin lines.}
 \label{fig_Vphi}
\end{figure*}

\begin{figure*}[htb]
\begin{center}
\includegraphics[width=6cm,height=4cm]{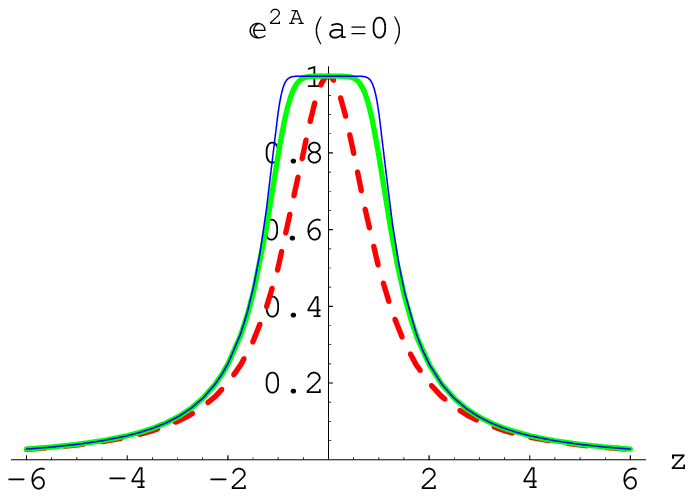}
\includegraphics[width=6cm,height=4cm]{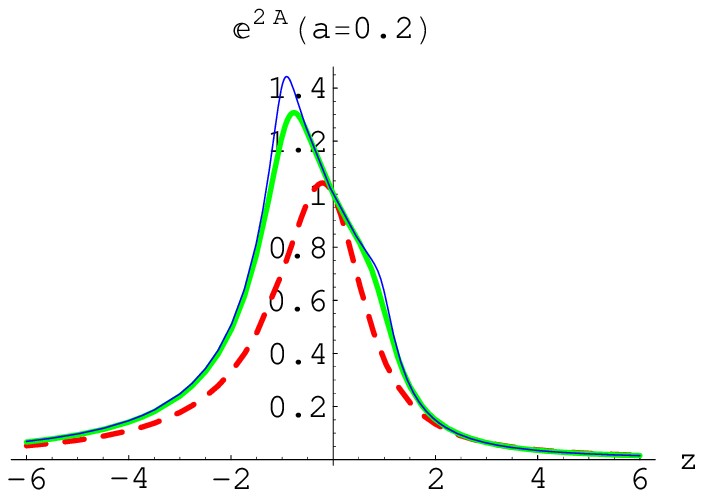}
\vskip 2mm
\includegraphics[width=6cm,height=4cm]{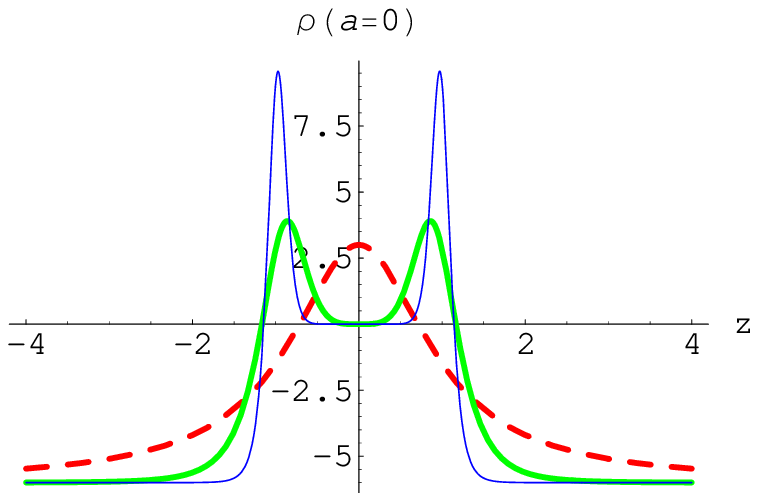}
\includegraphics[width=6cm,height=4cm]{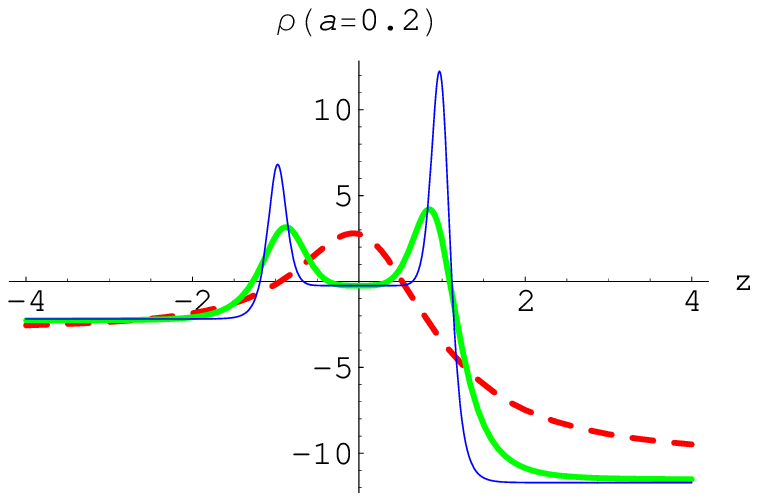}
\end{center}
\vskip -4mm \caption{(Color online) The shapes of the warp factor
$e^{2A}$ and the energy density $\rho$ for the symmetric ($a=0$) and
asymmetric ($a=0.2$) thick branes. The parameters are set to
$\lambda=1$, $s=1$ for red dashed lines, $s=3$ for green thick
lines, and $s=7$ for blue thin lines.}
 \label{fig_BPSBrane}
\end{figure*}

Based on the symmetric solution above, and in concordance with the
approach presented in \cite{asymdSBrane2}, a three-parameter family
of asymmetric thick branes in five dimensions were constructed in
Ref. \cite{asymdSBrane2}:
\begin{eqnarray}
 e^{2A}&=& \frac{1}{\left(1+(\lambda z)^{2s}\right)^{1/s}
        {\cal F}(z)^{2} },\qquad
      \label{e2A4} \\
\phi~~ &=&\phi_{0}\arctan(\lambda z)^s,
       \label{phi4} \\
V(\phi)&=&
      -\frac{3}{4}\sin^2(\phi/\phi_0)\tan^{-2/s}(\phi/\phi_0)
      {\cal K}(\phi)\nonumber \\
 &\times&  \bigg\{16a\tan^{1/s}(\phi/\phi_0)
      +\cos^{-2/s}(\phi/\phi_0)
     \bigg[5  \nonumber \\
 &&  ~~-2s-(3+2s)\cos(2\phi/\phi_0)\bigg]
     {\cal K}(\phi)\bigg\} \nonumber \\
 &-& 6a^2\cos^{2/s}(\phi/\phi_0),~~~~~~ \label{Vphi4}
\end{eqnarray}
where the asymmetric factor $a$ satisfies
\begin{equation}
 0<a <\frac{\Gamma(1/s)\;\lambda}{\Gamma(1/2s)\ \Gamma(1+1/2s)}
 ~\left(>\frac{\lambda}{2}\right), \label{constraintOna}
\end{equation}
${\cal F}(z)$ and  ${\cal K}(\phi)$ are defined as
\begin{eqnarray}
 {\cal F}(z)&\equiv& 1+a z\
       {}_2F_1\left(\frac{1}{2s},\frac{1}{s},1+\frac{1}{2s}
                    ,-(\lambda z)^{2s}\right), \label{calFz}\\
 {\cal K}(\phi)&\equiv&
   \lambda+a\tan^{1/s}\big({\phi}/{\phi_0}\big)\nonumber \\
      &&\times ~{}_2F_1\left(\frac{1}{2s},\frac{1}{s},1+\frac{1}{2s}
              ,-\tan^{2}\big(\frac{\phi}{\phi_0}\big)\right).
\end{eqnarray}
The parameter $a$ describes the asymmetry of the solution. For
$a\rightarrow 0$ and $s=1$ the regularized version of the
Randall-Sundrum thin brane will be recovered
\cite{GremmPLB2000,Guerrero2002}. For $\alpha>0$ and $s>1$, this is
a solution of an asymmetric static double domain wall space-time
interpolating between different AdS$_{5}$ vacua. The scalar
curvature $R$ and the energy density $\rho$ for the solution are
read
\begin{eqnarray}
 R &=& -\frac{40 a (\lambda z)^{2 s}}
             {z \left(1+(\lambda z)^{2s}\right)}{\cal F}(z)
     -\frac{20 a^2 }{\left(1+(\lambda z)^{2 s}\right)^{1/s}}
     \nonumber \\    &&
     -\frac{4 (\lambda z)^{2 s}
           \left(2-4 s+5 (\lambda z)^{2s}\right)}
         {z^2\left(1+(\lambda z)^{2 s}\right)^{2-{1}/{s}}} {\cal F}^2(z),~~~
              \label{R4} \\
 \rho &=&  -\frac{12 a  (\lambda z)^{2 s}}
                 {z \left(1+(\lambda z)^{2 s}\right)}{\cal F}(z)
     -\frac{6 a^2}{\left(1+(\lambda z)^{2s}\right)^{1/s}}
     \nonumber \\    &&
     -\frac{3  (\lambda z)^{2 s}
             \left(1-2 s+2 (\lambda z)^{2 s}\right)}
         {z^2\left(1+(\lambda z)^{2
         s}\right)^{2-{1}/{s}}}{\cal F}^2(z).~~~~~~~
       \label{EnergyDensity2}
\end{eqnarray}
The shapes of the kink $\phi$, the potential $V(\phi)$, the warp
factor $e^{2A}$, and the energy density $\rho$ are shown in Figs.
\ref{fig_Vphi} and \ref{fig_BPSBrane}. It is clear that, the
single brane is localized at $z=0$, while the two sub-branes are
localized at $z=\pm 1/\lambda$ and the thickness of the double
brane is $2/\lambda$. When $s\rightarrow\infty$, each sub-brane is
a thin brane. More detailed discussions can be found in Ref.
\cite{asymdSBrane2}.

\section{Localization and mass spectra of fermions on the thick branes}
\label{SecLocalize}

In this section let us investigate the localization problem of spin
1/2 fermions on the family of symmetric and asymmetric thick branes
given in Sec. \ref{SecModel} by means of the gravitational
interaction and sclar-fermion couplings. We will analyze the spectra
of fermions on the thick branes by present the potential of the
corresponding Schr\"{o}dinger equation. It can be seen from the
following calculations that the mass-independent potential can be
obtained conveniently with the conformally flat metric
(\ref{linee}).

In five dimensions, fermions are four component spinors and their
Dirac structure is described by $\Gamma^M= e^M _{\bar{M}}
\Gamma^{\bar{M}}$ with $\{\Gamma^M,\Gamma^N\}=2g^{MN}$, where
$\bar{M}, \bar{N}, \cdots =0,1,2,3,5$ denote the five-dimensional
local Lorentz indices, and $\Gamma^{\bar{M}}$ are the flat gamma
matrices in five dimensions. In our set-up,
$\Gamma^M=(\text{e}^{-A}\gamma^{\mu},\text{e}^{-A}\gamma^5)$,
where $\gamma^{\mu}$ and $\gamma^5$ are the usual flat gamma
matrices in the Dirac representation. The Dirac action of a
massless spin 1/2 fermion coupled to the scalar is
\begin{eqnarray}
S_{1/2} = \int d^5 x \sqrt{-g} \left(\bar{\Psi} \Gamma^M D_M
\Psi-\eta \bar{\Psi} F(\phi) \Psi\right), \label{DiracAction}
\end{eqnarray}
where the covariant derivative $D_M$ is defined as $D_M\Psi =
(\partial_M + \omega_M) \Psi$ with the spin connection $\omega_M=
\frac{1}{4} \omega_M^{\bar{M} \bar{N}} \Gamma_{\bar{M}}
\Gamma_{\bar{N}}$. With the metric (\ref{linee}), the nonvanishing
components of the spin connection $\omega_M$ are
\begin{eqnarray}
  \omega_\mu =\frac{1}{2}(\partial_{z}A) \gamma_\mu \gamma_5. \label{eq4}
\end{eqnarray}
Then the five-dimensional Dirac equation is read as
\begin{eqnarray}
 \left\{ \gamma^{\mu}\partial_{\mu}
         + \gamma^5 \left(\partial_z  +2 \partial_{z} A \right)
         -\eta\; \text{e}^A F(\phi)
 \right \} \Psi =0, \label{DiracEq1}
\end{eqnarray}
where $\gamma^{\mu} \partial_{\mu}$ is the Dirac operator on the
brane. Note that the sign of the coupling $\eta$ of the spinor
$\Psi$ to the scalar $\phi$ is arbitrary and represents a coupling
either to kink or to antikink domain wall. For definiteness, we
shall consider in what follows only the case of a kink coupling,
and thus assume that $\eta>0$.

Now we study the above five-dimensional Dirac equation. Because of
the Dirac structure of the fifth gamma matrix $\gamma^5$, we
expect the left- and right-handed projections of the
four-dimensional part to behave differently. From the equation of
motion (\ref{DiracEq1}), we will search for the solutions of the
general chiral decomposition
\begin{equation}
 \Psi(x,z) = \text{e}^{-2A}\sum_n\bigg(\psi_{Ln}(x) f_{Ln}(z)
 +\psi_{Rn}(x) f_{Rn}(z)\bigg),
\end{equation}
where $\psi_{Ln}(x)=-\gamma^5 \psi_{Ln}(x)$ and
$\psi_{Rn}(x)=\gamma^5 \psi_{Rn}(x)$ are the left-handed and
right-handed components of a four-dimensional Dirac field,
respectively, the sum over $n$ can be both discrete and
continuous. Here, we assume that $\psi_{L}(x)$ and $\psi_{R}(x)$
satisfy the four-dimensional massive Dirac equations
$\gamma^{\mu}\partial_{\mu}\psi_{Ln}(x)=m_n\psi_{R_n}(x)$ and
$\gamma^{\mu}\partial_{\mu}\psi_{Rn}(x)=m_n\psi_{L_n}(x)$. Then
$\alpha_{L}(z)$ and $\alpha_{R}(z)$ satisfy the following coupled
equations
\begin{subequations}
\begin{eqnarray}
 \left[\partial_z + \eta\;\text{e}^A F(\phi) \right]f_{Ln}(z)
  &=&  ~~m_n f_{Rn}(z), \label{CoupleEq1a}  \\
 \left[\partial_z- \eta\;\text{e}^A F(\phi) \right]f_{Rn}(z)
  &=&  -m_n f_{Ln}(z). \label{CoupleEq1b}
\end{eqnarray}\label{CoupleEq1}
\end{subequations}
From the above coupled equations, we get the Schr\"{o}dinger-like
equations for the KK modes of the left and right chiral fermions
\begin{eqnarray}
  \big(-\partial^2_z + V_L(z) \big)f_{Ln}
            &=&\mu_n^2 f_{Ln},
   \label{SchEqLeftFermion}  \\
  \big(-\partial^2_z + V_R(z) \big)f_{Rn}
            &=&\mu_n^2 f_{Rn},
   \label{SchEqRightFermion}
\end{eqnarray}
where the effective potentials are given by
\begin{subequations}
\begin{eqnarray}
  V_L(z)&=& \big(\eta\;\text{e}^{A}   F(\phi)\big)^2
     - \partial_z\big(\eta\;\text{e}^{A}   F(\phi)\big), \label{VL}\\
  V_R(z)&=&   V_L(z)|_{\eta \rightarrow -\eta}. \label{VR}
\end{eqnarray}\label{Vfermion}
\end{subequations}
In order to obtain the standard four-dimensional action for the
massive chiral fermions:
\begin{eqnarray}
 S_{1/2} &=& \int d^5 x \sqrt{-g} ~\bar{\Psi}
     \left(  \Gamma^M (\partial_M+\omega_M)
     -\eta F(\phi)\right) \Psi  \nn \\
  &=& \sum_{n}\int d^4 x \left(~\bar{\psi}_{Rn}
      \gamma^{\mu}\partial_{\mu}\psi_{Rn}
        -~\bar{\psi}_{Rn}m_{n}\psi_{Ln} \right) \nn \\
  &+&\sum_{n}\int d^4 x \left(~\bar{\psi}_{Ln}
      \gamma^{\mu}\partial_{\mu}\psi_{Ln}
        -~\bar{\psi}_{Ln}m_{n}\psi_{Rn} \right)  \nn \\
  &=&\sum_{n}\int d^4 x
    ~\bar{\psi}_{n}
      (\gamma^{\mu}\partial_{\mu} -m_{n})\psi_{n},
\end{eqnarray}
we need the following orthonormality conditions for $f_{L_{n}}$ and
$f_{R_{n}}$:
\begin{eqnarray}
&& \int_{-\infty}^{\infty} f_{Lm}(z) f_{Ln}(z)dz
  =\delta_{mn},\nonumber\\
&&\int_{-\infty}^{\infty} f_{Rm}(z) f_{Rn}(z)dz
  =\delta_{mn},\label{orthonormality}\\
&& \int_{-\infty}^{\infty} f_{Lm}(z) f_{Rn}(z)dz=0.
 \nonumber
\end{eqnarray}
Note that the differential equations (\ref{SchEqLeftFermion}) and
(\ref{SchEqRightFermion}) can be factorized as
\begin{eqnarray}
 \left[-\partial_z+\eta\;\text{e}^A F(\phi)\right]
 \left[\partial_z+\eta\;\text{e}^A F(\phi) \right]
 f_{Ln}(z) &=& m_n^2 f_{Ln}(z), \label{SchEqLeftFermion2} \nonumber
 \\ \\
 \left[-\partial_z-\eta\;\text{e}^A F(\phi)\right]
 \left[\partial_z-\eta\;\text{e}^A F(\phi) \right]
 f_{Rn}(z) &=& m_n^2 f_{Rn}(z). \label{SchEqRightFermion2}\nonumber
 \\
\end{eqnarray}
It can be shown that $m_n^2$ is zero or positive since the resulting
Hamiltonian can be factorized as the product of two operators which
are adjoints of each other. Hence the system is stable against
linear classical metric and scalar fluctuations.

It can be seen that, in order to localize left or right chiral
fermions, there must be some kind of scalar-fermion coupling, and
the effective potential $V_L(z)$ or $V_R(z)$ should have a minimum
at the location of the brane. Furthermore, for the kink
configuration of the scalar $\phi(z)$ (\ref{phi3}), $F(\phi(z))$
should be an odd function of $\phi(z)$ when one demands that
$V_{L,R}(z)$ are invariant under $Z_2$ reflection symmetry
$z\rightarrow -z$. Thus we have $F(\phi(0))=0$ and
$V_L(0)=-V_R(0)=-\eta\partial_z F(\phi(0))$, which results in the
well-known conclusion: only one of the massless left and right
chiral fermions could be localized on the brane. The spectra are
determined by the behavior of the potentials at infinity. For
$V_{L,R}\rightarrow 0$ as $|z|\rightarrow \infty$, one of the
potentials would have a volcanolike shape and there exists only a
bound massless mode followed by a continuous gapless spectrum of
KK states, while another could not trap any bound states and the
spectrum is also continuous and gapless. The simplest Yukawa
coupling $F(\phi)=\phi$ and the generalized coupling
$F(\phi)=\phi^k$ with positive odd integer $k\;(\geq3)$ belong to
this type. For $V_{L,R}\rightarrow V_{\infty}=$ constant as
$|z|\rightarrow \infty$, those modes with $m_n^2<V_{\infty}$
belong to discrete spectrum and modes with $m_n^2>V_{\infty}$
contribute to a continuous one. For this case, the simplest
coupling is of the form $F(\phi)=\tan^{1/s}(\phi/\phi_0)$.  If the
potentials increase as $|z|\rightarrow \infty$, the spectrum is
discrete. There are a lot of couplings for the case. The concrete
behavior of the potentials is dependent on the function $F(\phi)$.
In what following, we will discuss in detail three typical
couplings for the above three cases as examples.

\subsection{Case I: $F(\phi)=\phi^k$}
\label{sec3.1}

We mainly consider the simplest case $F(\phi)=\phi$, for which the
explicit forms of the potentials (\ref{Vfermion}) are
\begin{eqnarray}
 V^S_L(z) &=& 3\eta^2\frac{(2s-1)}{s^2}
           \frac{\arctan^2(\lambda^s z^s)}
           {[1+({\lambda}z)^{2s}]^{\frac{1}{s}}}
           \nonumber \\
     &-& \eta\frac{\sqrt{6s-3}}{s}
        \frac{({\lambda}z)^{s}\left[s -({\lambda}z)^{s}
                \arctan(\lambda^s z^s)\right]}
             {z[1+({\lambda}z)^{2s}]^{1+\frac{1}{2s}}}
      , ~~~~~ \label{VSL_CaseI} \\
  V^S_R(z) &=& V^S_L(z)|_{\eta \rightarrow -\eta},
  \label{VSR_CaseI}
\end{eqnarray}
and
\begin{eqnarray}
 V^A_L(z)
 &=& \bigg\{3\eta^2\frac{(2s-1)}{s^2}
           \frac{\arctan^2(\lambda^s z^s)}
           {[1+({\lambda}z)^{2s}]^{\frac{1}{s}}}
           \nonumber \\&&
    ~~ + a \eta \frac{\sqrt{6s-3}}{s}
        \frac{\arctan(\lambda^s z^s)}
          {\left[1+(z\lambda)^{2s}\right]^{\frac{3}{2s}}}
     \bigg\}  \frac{1}{{\cal F}^2(z)}           \\
 &&- \eta\frac{\sqrt{6s-3}}{s}
        \frac{({\lambda}z)^{s}\left[s -({\lambda}z)^{s}
                \arctan(\lambda^s z^s)\right]}
             {z[1+({\lambda}z)^{2s}]^{1+\frac{1}{2s}}
               {{\cal F}(z)} }
        ,~~~~\nonumber \\
  V^A_R(z) &=& V^A_L(z)|_{\eta \rightarrow -\eta},
\end{eqnarray}
for the symmetric and asymmetric brane solutions, respectively.

\begin{figure*}[htb]
\begin{center}
\includegraphics[width=6cm,height=4cm]{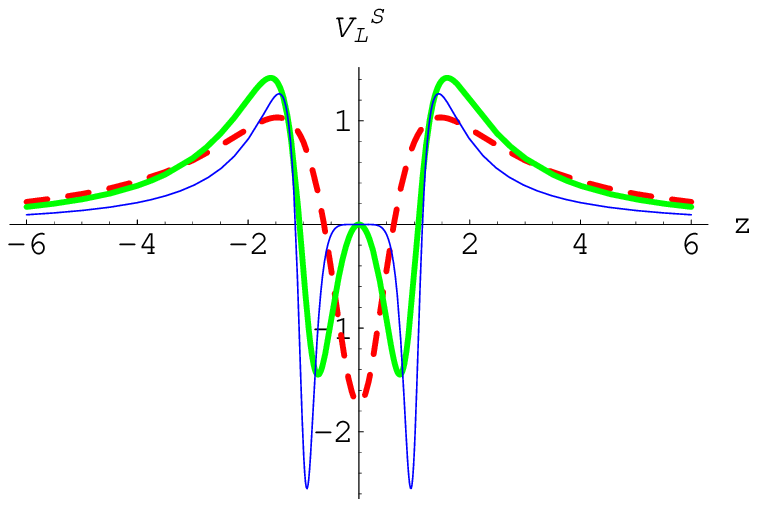}
\includegraphics[width=6cm,height=4cm]{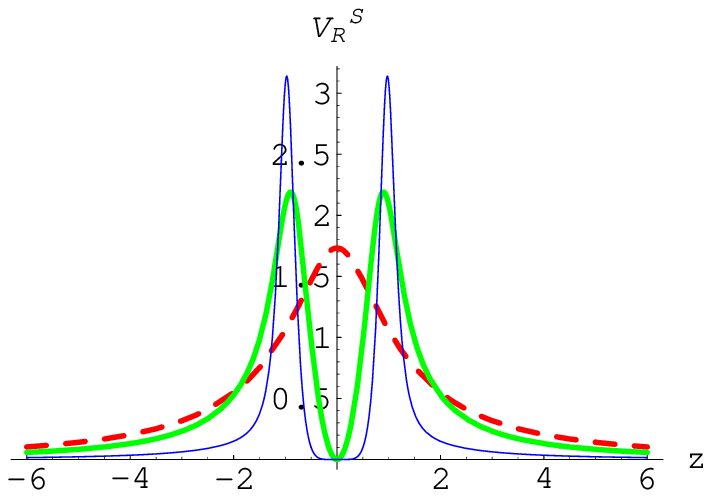}
\vskip 4mm
\includegraphics[width=6cm,height=4cm]{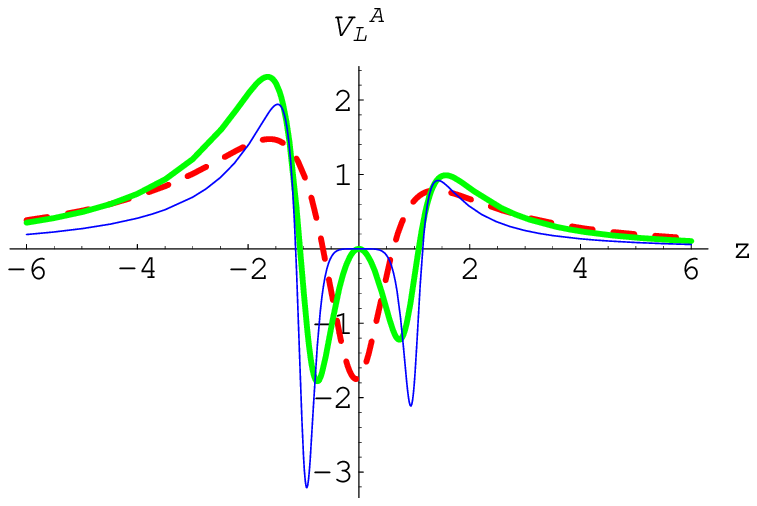}
\includegraphics[width=6cm,height=4cm]{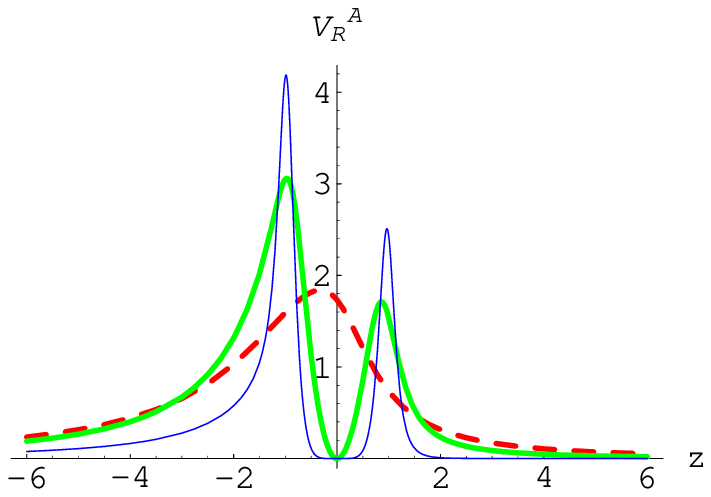}
\end{center}
\vskip -4mm \caption{(Color online) The shapes of the symmetric
potentials $V^S_{L,R}(a=0)$ and the asymmetric potentials
$V^A_{L,R}(a=0.2)$ for left and right chiral fermions for the case
$F(\phi)=\phi$. The parameters are set to $\eta=\lambda=1$ and $s=1$
for red dashed lines, $s=3$ for green thick lines, and $s=7$ for
blue thin lines. }
 \label{fig_VFermion_Case1}
\end{figure*}

\begin{figure*}[htb]
\begin{center}
\includegraphics[width=6cm,height=4cm]{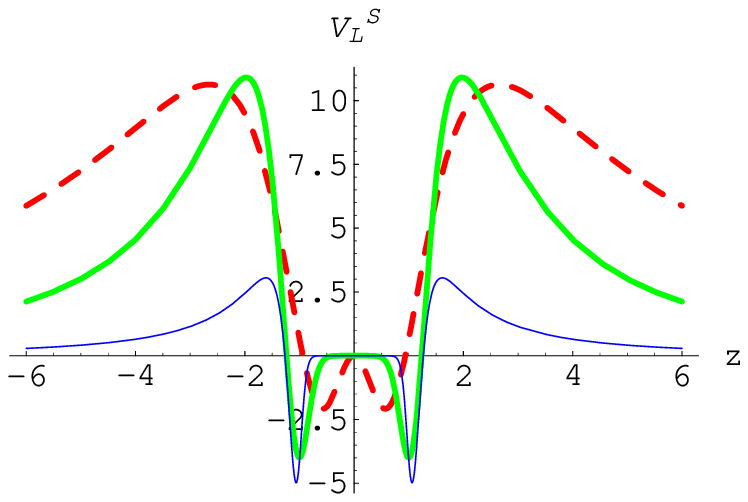}
\includegraphics[width=6cm,height=4cm]{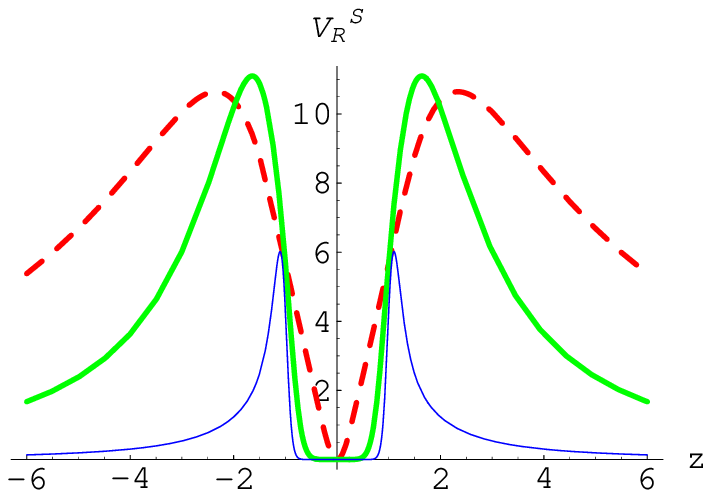}
\end{center}
\vskip -4mm \caption{(Color online) The shapes of the potentials
$V^S_{L,R}$ for left and right chiral fermions for the case
$F(\phi)=\phi^3$. The parameters are set to $\eta=\lambda=1$, $s=1$
for red dashed lines, $s=3$ for green thick lines, and $s=7$ for
blue thin lines. }
 \label{fig_VFermion_Case1b}
\end{figure*}

All potentials have the asymptotic behavior:
$V^{S,A}_{L,R}(z\rightarrow\pm \infty)\rightarrow0$. The values of
the potentials for left and right chiral fermions at $z = 0$ are
given by
\begin{equation}
 V^{S,A}_L(0) =-V^{S,A}_R(0)
 = \left\{
\begin{array}{c}
  -\sqrt{3}\;\eta\lambda \\
  0
\end{array}
\begin{array}{l}
  ~~~\text{for} ~~~ s=1. \\
  ~~~\text{for} ~~~ s>1.
\end{array} \right.
\end{equation}
So for a given coupling constant $\eta$ and $\lambda$, the values of
the potentials for left and right chiral fermions at $z=0$ are
opposite for $s=1$ and vanish for $s>1$. Note that there are a
single brane and a double brane for  $s=1$ and $s>1$, respectively.
The shapes of the potentials are shown in Fig.
\ref{fig_VFermion_Case1} for given values of positive $\eta$ and
$\lambda$. It can be seen that $V_L(z)$ is indeed a modified volcano
type potential for the single brane scenario with $s=1$, and it has
a well. While for the double brane case with $s>1$, the
corresponding potential $V_L(z)$ has a double well, and the
potential $V_R(z)$ of right chiral fermions has a single ``well",
which indicates that there may exist resonances (qusi-localized KK
modes). The shape of the potentials is relative to the inner
structure of the brane, or equivalently, it depends partly on the
warp factor $e^{2A}$. Furthermore, the coupling type of scalar and
fermion also affects the structure of the potentials. For example,
for the case $F(\phi)=\phi^k$ with positive odd integer $k\geq 3$,
we have $V^{S,A}_{L,R}(0)=0$ for both $s=1$ and $s>1$, and the
potentials for left and right chiral fermions have a double well and
a single well even for $s=1$, respectively (see Fig.
\ref{fig_VFermion_Case1b}).

Since $V^{S,A}_L(z)\rightarrow 0$ when $z\rightarrow\pm\infty$, the
potentials for left chiral fermions provides no mass gap to separate
the fermion zero mode from the excited KK modes. Because the
potentials $V^{S,A}_R(z)\geq 0$, there is no bound right chiral
fermion zero mode. For both left and right chiral fermions, there
exists a continuous gapless spectrum of the KK modes.

For positive $\eta$, only the potentials for left chiral fermions
have a negative single well and a double well at the location of the
branes for single brane and double brane, respectively, which could
trap the left chiral fermion zero mode solved from
(\ref{CoupleEq1a}) by setting $m_0=0$:
\begin{equation}
 f_{L0}(z)
 \propto \exp\left(-\eta\int^z d\bar{z}\text{e}^{A(\bar{z})}\phi(\bar{z})\right).
  \label{zeroModefL0CaseI}
\end{equation}
In order to check the normalization condition (\ref{orthonormality})
for the zero mode (\ref{zeroModefL0CaseI}), we need to check whether
the inequality
\begin{equation}
 \int f_{L0}^2(z) dz \propto
 \int \exp\left(-2\eta\int^z d\bar{z} \text{e}^{A(\bar{z})}\phi(\bar{z})\right) dz
  < \infty     \label{condition1}
\end{equation}
is satisfied. For the integral $\int dz\text{e}^{A}\phi$, we only
need to consider the asymptotic characteristic of the function
$\eta\;\text{e}^{A}\phi$ for $z \rightarrow \infty$. For the
asymmetric brane scenario, we have
\begin{eqnarray}
 2\;\text{e}^{A} \phi
 &=& \frac{2{\sqrt{3(2s-1)}}\arctan(\lambda z)^s
          \left(1+(\lambda z)^{2s}\right)^{-1/2s}}
        {{s}\left[1+a z\
          {}_2F_1\left(\frac{1}{2s},\frac{1}{s},1+\frac{1}{2s}
                       ,-(\lambda z)^{2s}\right)\right]}
   \nonumber \\
 &\rightarrow& \frac{1}{{\eta_0}z} ~~~~~ \text{for} ~~~z \rightarrow \infty,
\end{eqnarray}
where the constant ${\eta_0}$ is given by
\begin{eqnarray}
 {\eta_0} = \frac{s\lambda }{{\sqrt{3(2s-1)}}\;\pi}
         \left( 1+ a\frac{\Gamma(1+1/2s)\Gamma(1/2s)}
                {\lambda\;\Gamma(1/s)}   \right).
\end{eqnarray}
For the symmetric brane case we only need to take $a=0$. So, when $z
\rightarrow \infty$, we have
\begin{eqnarray}
f_{L0}^2(z)\propto\exp\left(-2\eta\int^z
d\bar{z}\text{e}^{A(\bar{z})}\phi(\bar{z})\right)
    \rightarrow z^{-\eta/{\eta_0}},
\end{eqnarray}
which indicates that the normalization condition (\ref{condition1})
is
\begin{equation}
 \eta>{\eta_0}.    \label{conditionCaseI}
\end{equation}
Provided the condition (\ref{conditionCaseI}), the zero mode of
left chiral fermions can be localized on the brane. In Refs.
\cite{KoleyCQG2005,LiuPRD2008}, it was shown that the
corresponding zero mode can also be localized on the brane in the
background of Sine-Gordon kinks provided similar condition as
(\ref{conditionCaseI}). While the fermion zero mode can not be
localized on the de Sitter brane with the same coupling
$F(\phi)=\phi$ \cite{LiuJCAP2009}. Note that for large $s$,
${\eta_0}$ can be approximated as
\begin{eqnarray}
 {\eta_0} \approx \frac{(\lambda+2a)}{{\sqrt{6}}\;\pi}
               \sqrt{s}.
\end{eqnarray}
It is clear that in order for the potentials to localize the zero
mode of left chiral fermions for larger $s$, $\lambda$ or
asymmetric factor $a$, the stronger coupling of kink and fermions
is required. That is to say, the massless mode of left chiral
fermion is most easy to be localized on the symmetric single
brane. The asymmetric factor $a$ may destroy the localization of
massless fermions. This is different from the situation of the
zero modes of scalars and vectors on symmetric and asymmetric de
Sitter branes \cite{LiuJCAP2009} , where increasing the asymmetric
factor $a$ does not change the number of the bound vector KK modes
but would increase that of the bound scalar KK modes, and the zero
modes of scalars and vectors are always localized on the de Sitter
branes.

\begin{figure*}[htb]
\begin{center}
\includegraphics[width=6cm,height=4cm]{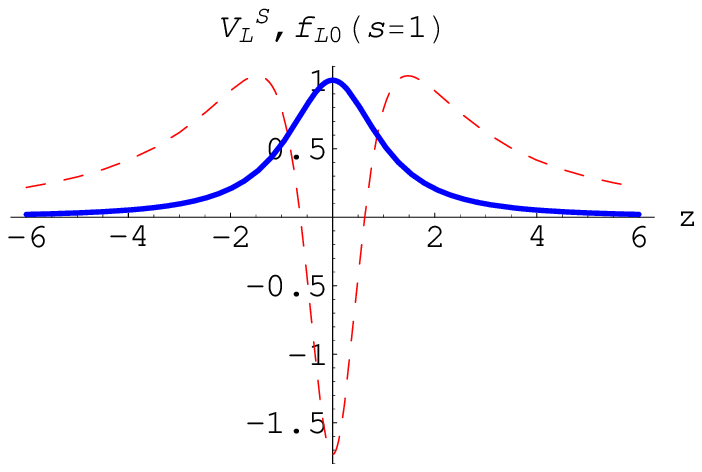}
\includegraphics[width=6cm,height=4cm]{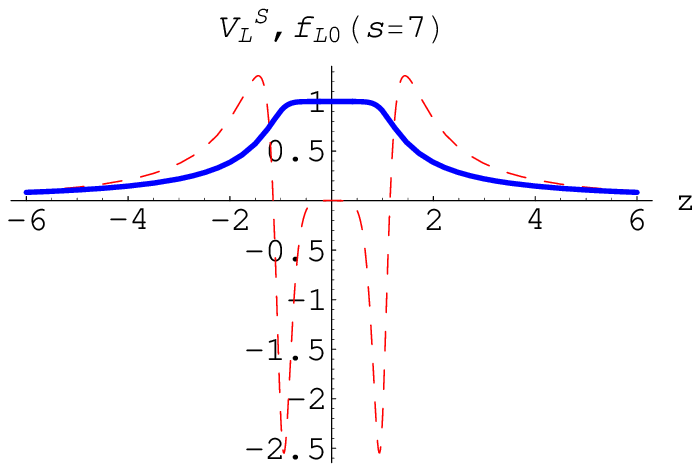}
\vskip 4mm
\includegraphics[width=6cm,height=4cm]{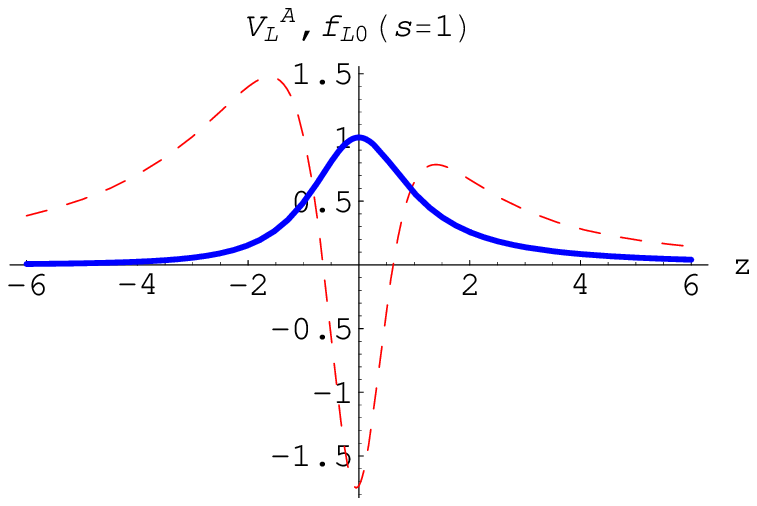}
\includegraphics[width=6cm,height=4cm]{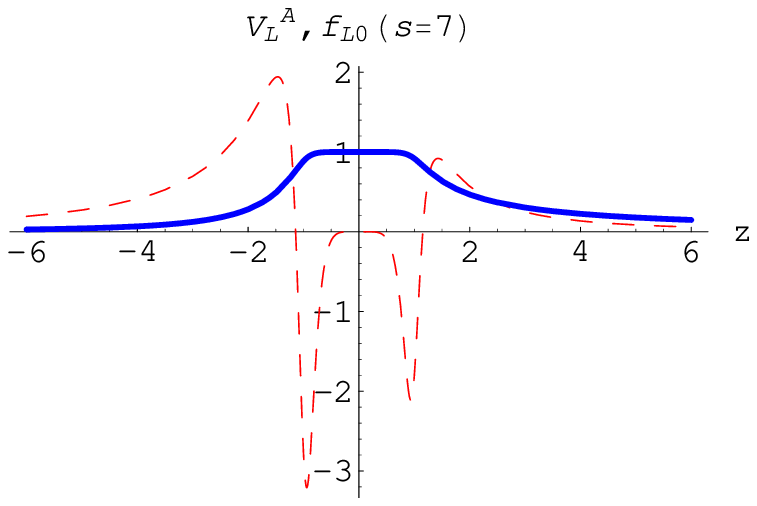}
\end{center}
\vskip -4mm \caption{(Color online) The shapes of the zero modes
$f_{L0}(z)$ (blue thick lines)
 and the potentials  $V^{S,A}_L(z)$ (red dashing lines) for the symmetric
 and asymmetric thick branes for the case $F(\phi)=\phi$. The parameters are set to
 $\eta=\lambda=1$, $s=1$ and 7, and $a=0$ (up two) and $0.2$ (down two).}
 \label{fig_zeroModefL0CaseI}
\end{figure*}

In Fig. \ref{fig_zeroModefL0CaseI}, we plot the left chiral fermion
potentials $V^{S,A}_L(z)$ and the corresponding zero modes. We see
that the zero modes are bound on the brane. They represents the
lowest energy eigenfunction (ground state) of the Schr\"{o}dinger
equation (\ref{SchEqLeftFermion}) since they have no zeros. Since
the ground state has the lowest mass square $m_0^2=0$, there is no
tachyonic left chiral fermion mode. The zero mode on both the
symmetric and asymmetric double walls is essentially constant
between the two interfaces. This is very different from the case of
gravitons, scalars and vectors, where the massless modes on the
asymmetric double wall are strongly localized only on the interface
centered around the lower minimum of the potential.  The massive
modes will propagate along the extra dimension and those with lower
energy would experience an attenuation due to the presence of the
potential barriers near the location of the brane.

The potential $V_R$ is always positive near the brane location and
vanishes when far away from the brane. This shows that it could
not trap any bound fermions with right chirality and there is no
zero mode of right chiral fermions. However, the shape of the
potential is strongly dependent on the parameter $s$. For $s\geq
3$, a potential well around the brane location would appear and
the well becomes deeper and deeper with increase of $\eta$. The
appearance of the potential well could be related to resonances,
i.e., massive fermions with a finite lifetime
\cite{0901.3543,Liu0904.1785}. In Ref. \cite{0901.3543}, a similar
potential and resonances for left and right chiral fermions were
found in background of two-field generated thick branes with
internal structure. We can investigate the massive modes of
fermions by solving numerically Eqs. (\ref{SchEqLeftFermion}) and
(\ref{SchEqRightFermion}).

\begin{figure*}[htb]
\begin{center}
\includegraphics[width=6cm,height=4cm]{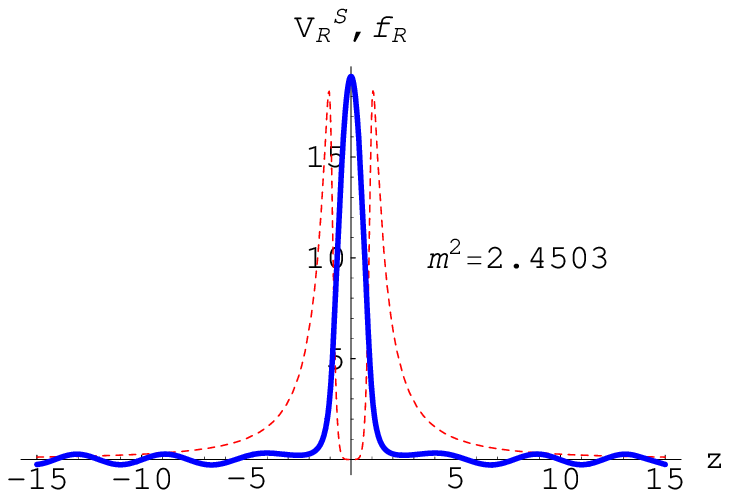}
\includegraphics[width=6cm,height=4cm]{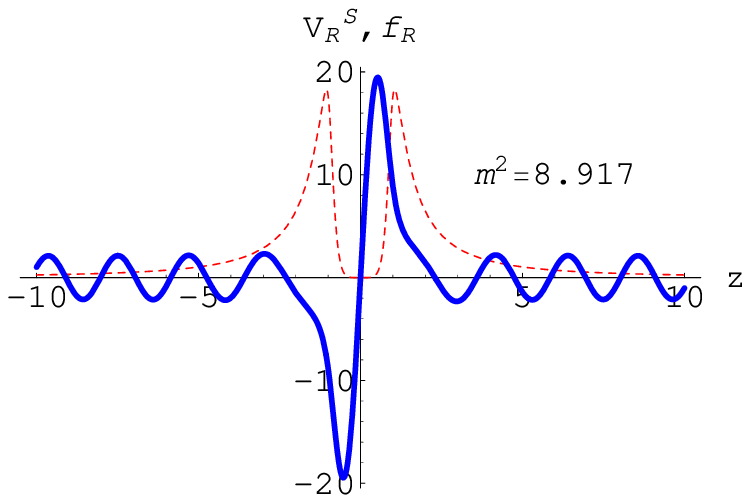}
\vskip 4mm
\includegraphics[width=6cm,height=4cm]{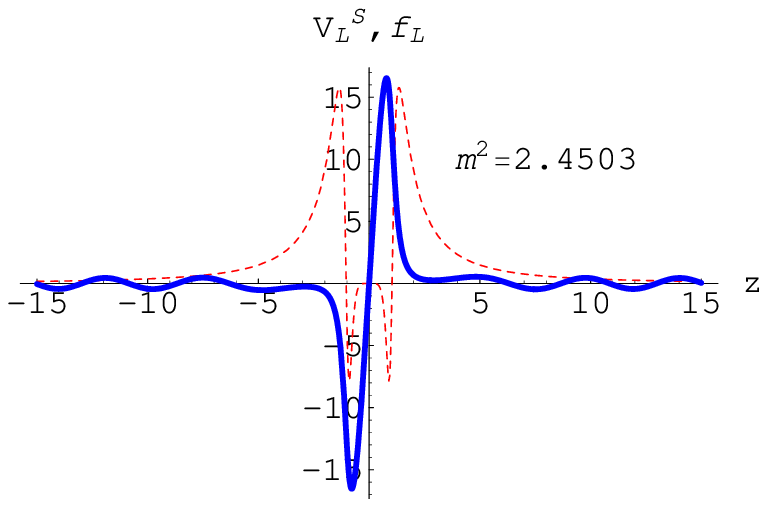}
\includegraphics[width=6cm,height=4cm]{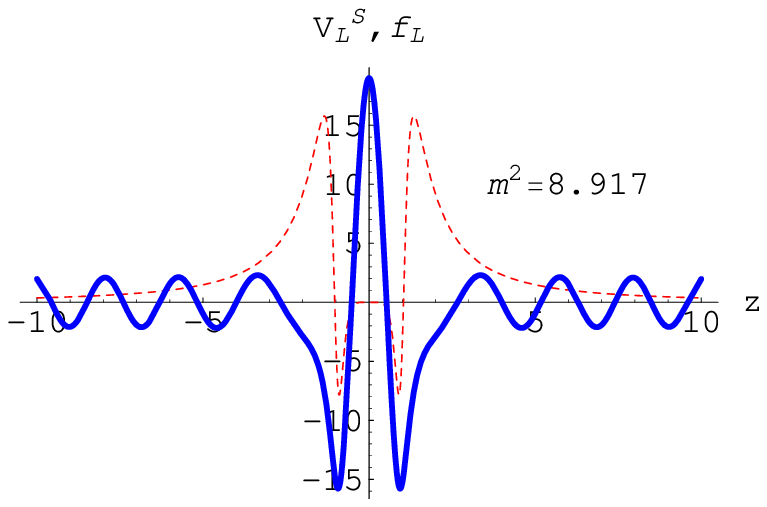}
\end{center}
\vskip -4mm \caption{(Color online) The shapes of the fermion
resonances (massive modes) $f_{L,R}(z)$ (blue thick lines)
 and the potentials  $V^{S}_{L,R}(z)$ (red dashing lines) for the symmetric
 thick branes for the case $F(\phi)=\phi$. The parameters are set to
 $\eta=4,\lambda=1$ and $s=7$.}
 \label{fig_fResonanceCase1}
\end{figure*}

\begin{figure*}[htb]
\begin{center}
\subfigure[$n=0,\Delta m=0.001177$] {\label{fig:PR1Case1}
\includegraphics[width=6cm,height=4cm]{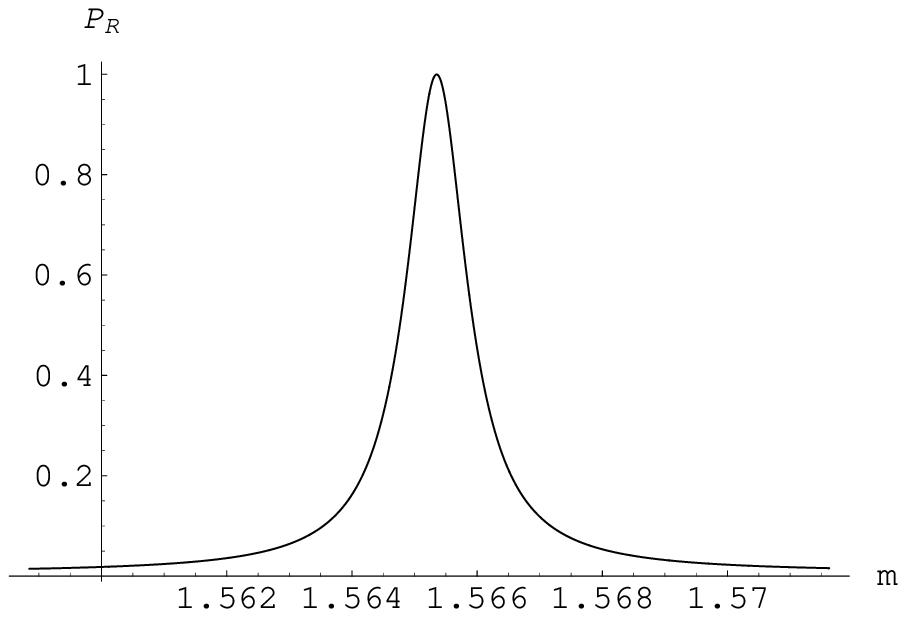}}
\subfigure[$n=1,\Delta m=0.04606$] {\label{fig:PR2Case1}
\includegraphics[width=6cm,height=4cm]{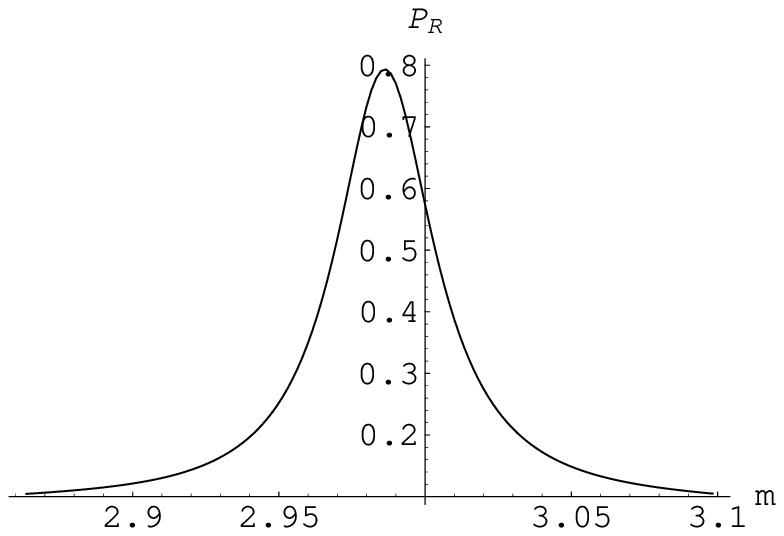}}
 \subfigure[$n=1,\Delta m=0.001184$]{\label{fig:PL1Case1}
\includegraphics[width=6cm,height=4cm]{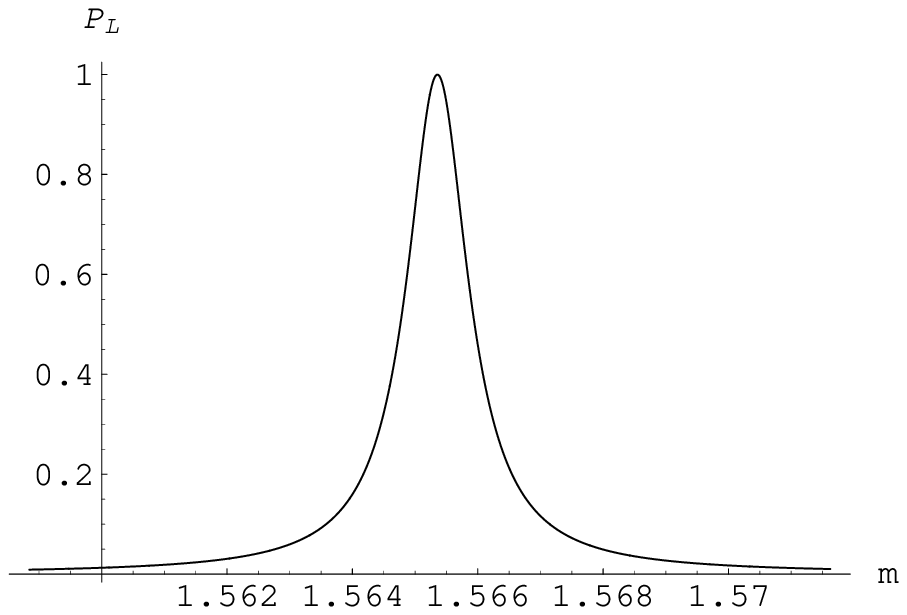}}
\subfigure[$n=2,\Delta m=0.04580$] {\label{fig:PL2Case1}
\includegraphics[width=6cm,height=4cm]{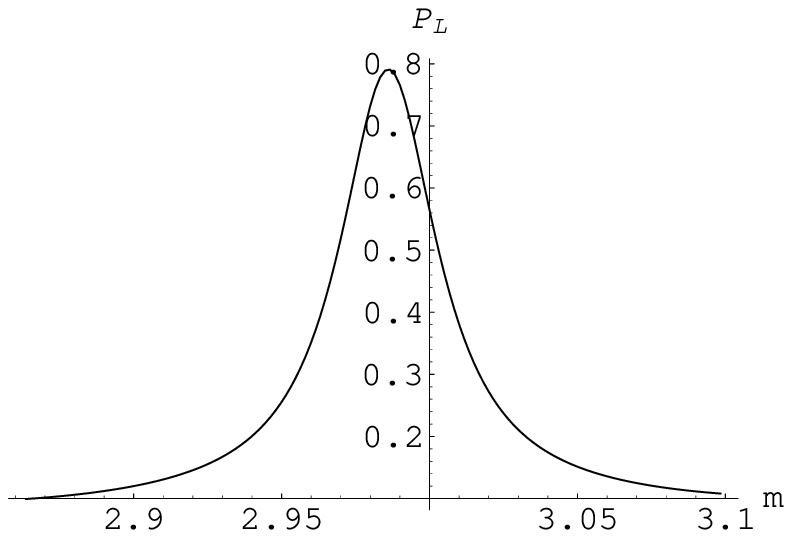}}
\end{center}
\vskip -4mm \caption{The probability for finding massive KK modes
of right and left chiral fermions around the brane location, $P_R$
and $P_L$, as functions of $m$ for the symmetric thick branes for
the case $F(\phi)=\phi$. The parameters are set to
 $\eta=4,\lambda=1$ and $s=7$.}
 \label{fig_PLPRResonanceCase1}
\end{figure*}

In Ref. \cite{0901.3543}, the authors suggested that large peaks
in the distribution of $f_{L,R}(0)$ as a function of $m$ would
reveal the existence of resonant states. In Ref.
\cite{Liu0904.1785}, we extended this idea and proposed that large
relative probabilities for finding massive KK modes within a
narrow range $-z_b<z<z_b$ around the brane location, are called
$P_{L,R}$, would indicate the existence of resonances. The
relative probabilities are defined as follows:
\begin{equation}
 P_{L,R}(m)=\frac{\int_{-z_b}^{z_b} |f_{L,R}(z)|^2 dz}
                 {\int_{-z_{max}}^{z_{max}} |f_{L,R}(z)|^2 dz},
 \label{Probability}
\end{equation}
where we choose $z_b=0.1 z_{max}$. For the set of parameters:
$\eta=4,\lambda=1$ and $s=7$, we find two resonances with mass
square 2.4503 and 8.917 for both left and right chiral fermions
(see Fig. \ref{fig_fResonanceCase1}). The configurations of Figs.
\ref{fig_fResonanceCase1}c and \ref{fig_fResonanceCase1}d could
present the $n=1$ and $n=2$ level KK resonance modes of left
chiral fermions. The $n=0$ level mode with left chirality is in
fact the only one bound state (the zero mode). While the
configurations of Figs. \ref{fig_fResonanceCase1}a and
\ref{fig_fResonanceCase1}b present the $n=0$ and $n=1$ level
resonances of right chiral fermions. We note that the spectra of
massive left-handed and right-handed fermionic resonances are the
same, which demonstrates that a Dirac fermion could be composed
from the left and right resonance KK modes \cite{Liu0904.1785}.
The lifetime $\tau$ for a resonance can be estimated by the width
in mass $\Gamma=\Delta{m}$ at half maximum of the corresponding
peak in Fig. \ref{fig_PLPRResonanceCase1}, which means that the
fermion disappears into the extra dimension with time
$\tau\sim\Gamma^{-1}$ \cite{RubakovPRL2000}. The lifetime of the
resonances are listed in Table \ref{tab1}.

\begin{table}[h]
\begin{center}
\caption{The mass, width, and lifetime for resonances of left and
right chiral fermions. The parameters are $\delta=\beta=0.5$ and
$\eta=10$.}\label{tab1}
\renewcommand\arraystretch{1.3}
\begin{tabular}
 {|l||c|c|c|c|}
  \hline
  $~$ & $m^2$ & $m$ & $\Gamma$ & $\tau$  \\
  \hline  \hline
  $n=0$(right) & 2.4503 & 1.5653 & 0.001177 & 849.6 \\
  \hline
  $n=1$(left) & 2.4503 & 1.5653 & 0.001184 &  844.5  \\
  \hline   \hline
  $n=1$(right) & 8.9179 & 2.9863 & 0.04606 & 21.71    \\
  \hline
  $n=2$(left) & 8.9179  & 2.9863 & 0.04580 &  21.83   \\
  \hline
\end{tabular}
\end{center}
\end{table}

\subsection{Case II: $F(\phi)=\tan^{1/s}(\phi/\phi_0)$}
\label{sec3.2}

Next, we consider the case $F(\phi)=\tan^{1/s}(\phi/\phi_0)$, for
which the potentials take the forms of
\begin{eqnarray}
 V^S_L(z) &=& \frac{\eta^2({\lambda}z)^{2}}
                {[1+({\lambda}z)^{2s}]^{\frac{1}{s}}}
     - \frac{\eta\lambda}
            {[1+({\lambda}z)^{2s}]^{1+\frac{1}{2s}}}
      ,  \label{VSL_CaseII} \\
  V^S_R(z) &=& 
    \frac{\eta^2({\lambda}z)^{2}}
                {[1+({\lambda}z)^{2s}]^{\frac{1}{s}}}
     + \frac{\eta\lambda}
            {[1+({\lambda}z)^{2s}]^{1+\frac{1}{2s}}}
      ,  \label{VSR_CaseII}
\end{eqnarray}
and
\begin{eqnarray}
 V^A_L(z)
 &=& \left\{
       \frac{\eta^2 ({\lambda}z)^2 }
           { {\left[1+({\lambda}z)^{2s}\right]^{\frac{1}{s}}}  }
      +\frac{a \eta {\lambda}z    }
            { {\left[1+({\lambda}z)^{2s}\right]^{\frac{3}{2s}}} }
     \right\} \frac{1}{ {\cal{F}}^{2}(z) }
     \nonumber \\ &&
  -\frac{\eta\lambda}
        { [1+({\lambda}z)^{2s}]^{1+\frac{1}{2s}} }
   \frac{1}{ {\cal{F}}(z) }, \label{VAL_CaseII} \\
  V^A_R(z) &=& V^A_L(z)|_{\eta \rightarrow -\eta},\label{VAR_CaseII}
\end{eqnarray}
for the symmetric and asymmetric brane solutions, respectively.

\begin{figure*}[htb]
\vskip -5mm
\begin{center}
\subfigure[$\eta=0.1$]{\label{fig_VSL_CaseIIa}
\includegraphics[width=6cm,height=4cm]{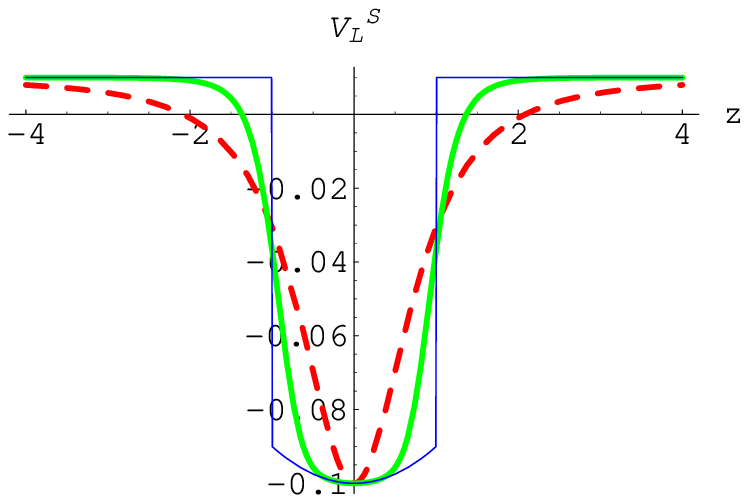}}
\subfigure[$\eta=0.1$]{\label{fig_VSR_CaseIIa}
\includegraphics[width=6cm,height=4cm]{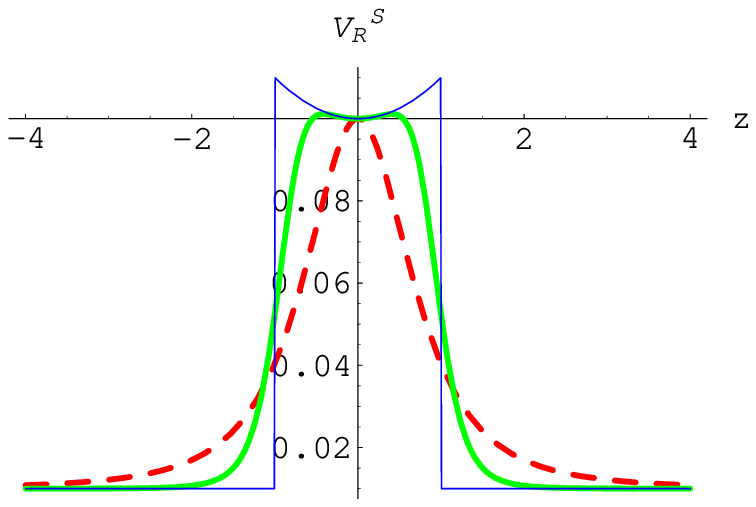}}
\subfigure[$\eta=1$]{\label{fig_VSL_CaseIIb}
\includegraphics[width=6cm,height=4cm]{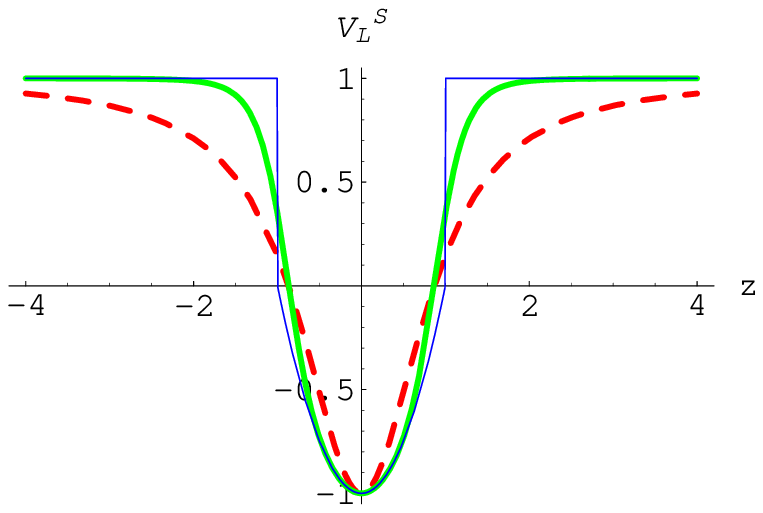}}
\subfigure[$\eta=1$]{\label{fig_VSR_CaseIIb}
\includegraphics[width=6cm,height=4cm]{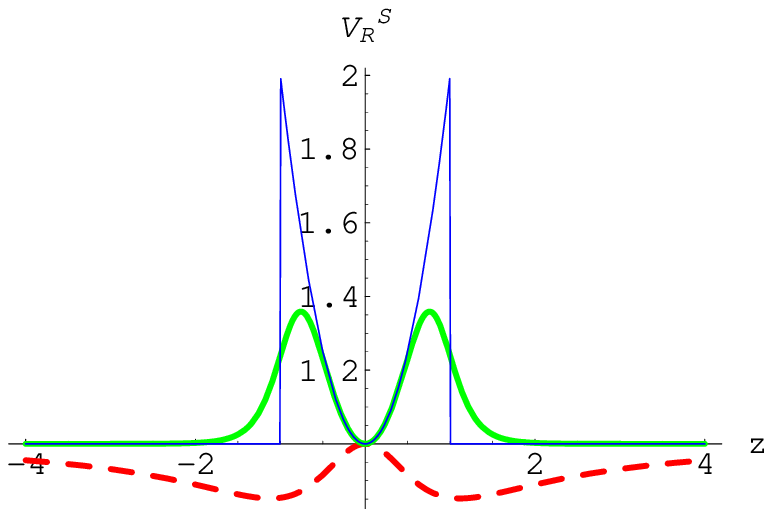}}
\subfigure[$\eta=6$]{\label{fig_VSL_CaseIIc}
\includegraphics[width=6cm,height=4cm]{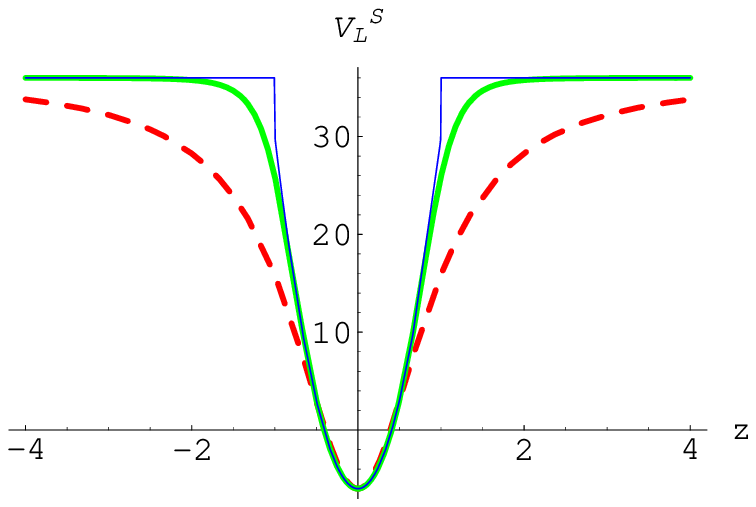}}
\subfigure[$\eta=6$]{\label{fig_VSR_CaseIIc}
\includegraphics[width=6cm,height=4cm]{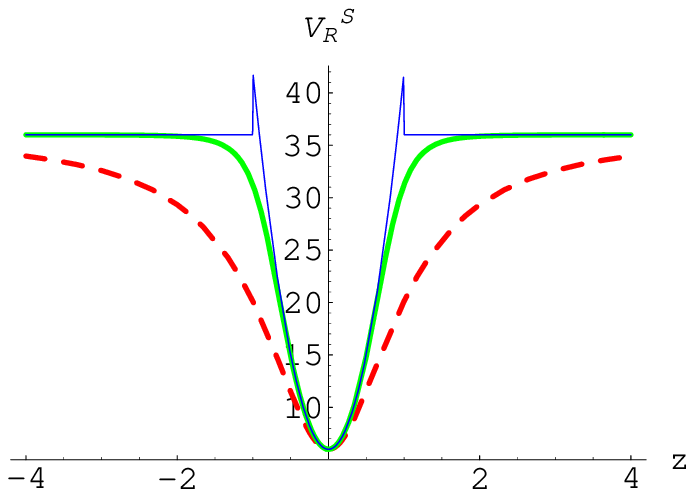}}
\end{center}
\vskip -7mm \caption{(Color online) The shapes of the potentials
$V^S_L(z)$ and $V^S_R(z)$ for the symmetric thick branes for the
case $F(\phi)=\tan^{1/s}(\phi/\phi_0)$. The parameter $s$ is set to
$s=1$ for red dashed lines, $s=3$ for green thick lines, and
$s\rightarrow \infty$ for blue thin lines. The parameter $\lambda$
is set to $\lambda=1$.}
 \label{fig_VSLR_CaseII}
\end{figure*}

~\\
\noindent\textbf{The symmetric potential}\\

\begin{figure*}[htb]
\begin{center}
\subfigure[$a=0,~s=1$ and $\infty $]{\label{fig_fL0_CaseII_s}
\includegraphics[width=6cm,height=4cm]{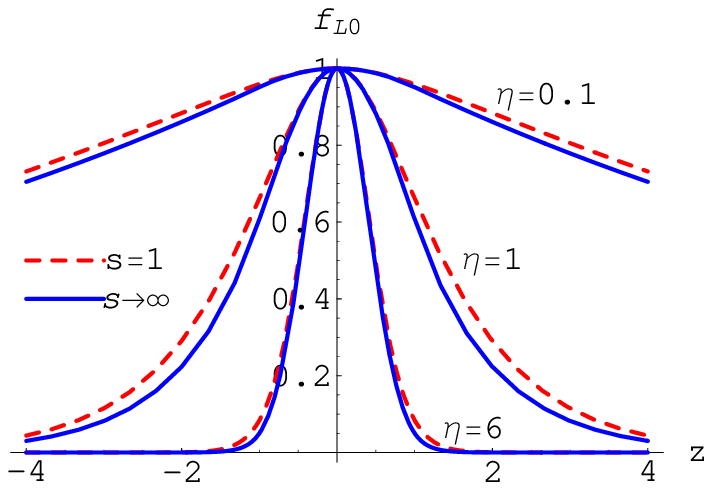}}
\subfigure[$s=1,~a=0$ and $0.4$]{\label{fig_fL0_CaseII_a}
\includegraphics[width=6cm,height=4cm]{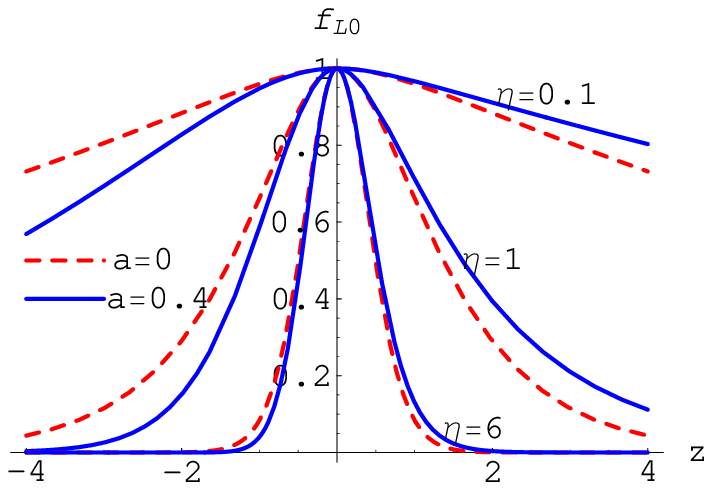}}
\end{center}
\vskip -4mm \caption{(Color online) The massless modes $f_{L0}(z)$
of left chiral fermions for the symmetric and asymmetric
 thick branes for the case $F(\phi)=\tan^{1/s}(\phi/\phi_0)$.
 The parameters are set to $\lambda=1$, $\eta=0.1$, $1$ and $6$.
 }
 \label{fig_fL0_CaseII}
\end{figure*}

We first investigate the potential $V^S_L(z)$ (\ref{VSL_CaseII}) for
the symmetric brane. It has a minimum (negative value)
$-\eta\lambda$ at $z=0$ and a maximum (positive value) $\eta^2$ at
$z=\pm\infty$. The shapes of the potential for various parameters
are plotted in Fig. \ref{fig_VSLR_CaseII}, from which we can see
that they are similar to that of a P\"{o}schl-Teller (PT) potential
for finite $s$. The massless KK mode can be solved as follows:
\begin{eqnarray}
 f_{L0}(z) \propto \exp\left\{-\frac{1}{2}\eta\lambda z^2
   {~_2F_1}\left(\frac{1}{s},\frac{1}{2s},
                 1+\frac{1}{s},-({\lambda}z)^{2s}\right)\right\}.
   \label{zeroModefSL0CaseII}
\end{eqnarray}
Because $f^2_{L0}(z)  \propto \exp(-2\eta z)$ when
$z\rightarrow\infty$, the massless KK mode is normalizable without
additional conditions, and it would be strongly localized on the
brane with large coupling constant $\eta$ (see Fig.
\ref{fig_fL0_CaseII_s}). We note that the potential $V^S_L(z)$
(\ref{VSL_CaseII}) and the zero mode $f_{L0}(z)$
(\ref{zeroModefSL0CaseII}) are very different from those given in
(\ref{VSL_CaseI}) and (\ref{zeroModefL0CaseI}) for left chiral
fermions:

(1) The potential (\ref{VSL_CaseII}) has a single well but the
potential (\ref{VSL_CaseI}) has a double well for $s\geq 3$, which
results in that the zero mode (\ref{zeroModefSL0CaseII}) is strongly
localized at the center of the double brane while the zero mode
(\ref{zeroModefL0CaseI}) is localized between the two sub-branes of
the double brane.

(2) The potential (\ref{VSL_CaseII}) tends to a positive constant
but the potential (\ref{VSL_CaseI}) runs to zero when far away from
the brane, which results in that the localization of the zero modes
here and (\ref{zeroModefL0CaseI}) is unconditional and conditional
(with condition (\ref{conditionCaseI})), respectively.

(3) The potential here provides mass gap to separate the zero mode
from the excited KK modes.

\begin{figure*}[htb]
\begin{center}
\subfigure[$\eta=1$]{\label{fig_fL_Spectra_CaseIIa}
\includegraphics[width=6cm,height=4cm]{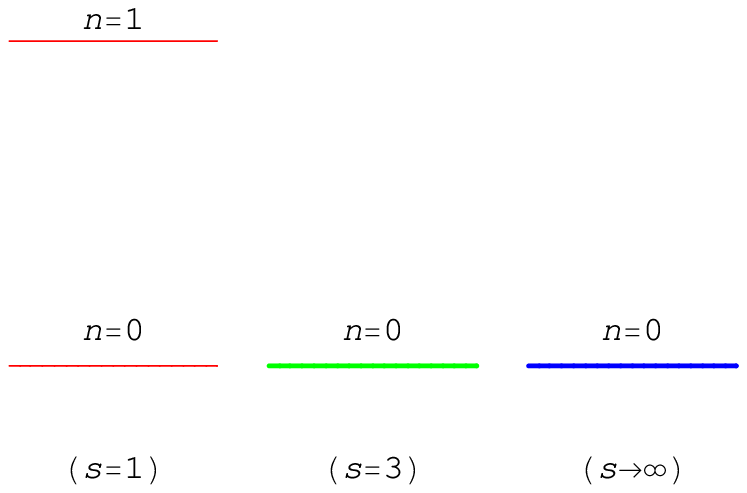}}
\hskip 5mm
 \subfigure[$\eta=6$]{\label{fig_fL_Spectra_CaseIIb}
\includegraphics[width=6cm,height=4cm]{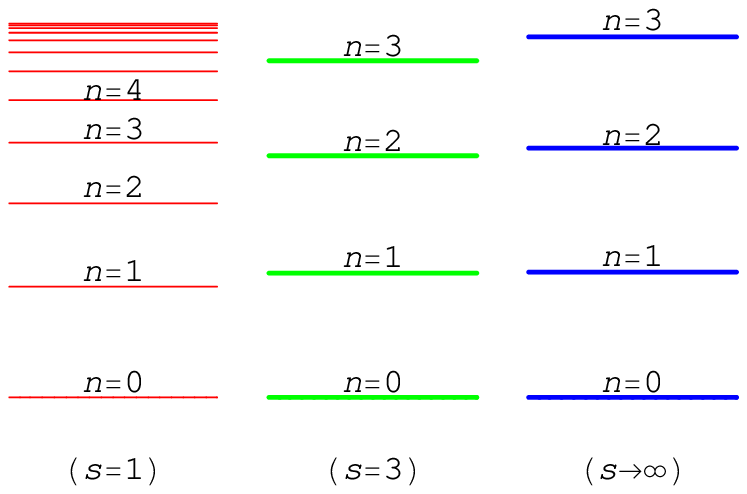}}
\end{center}
\vskip -4mm \caption{(Color online) The spectra $m_{Ln}^2$ of bound
KK modes of left chiral fermions for the symmetric thick branes with
$F(\phi)=\tan^{1/s}(\phi/\phi_0)$.
 The parameters are set to $\lambda=1$, $\eta=1$ and $6$, $s=1,3$
 and $s\rightarrow\infty$.}
 \label{fig_fSL_Spectra_CaseII}
\end{figure*}

We have known that the massless KK mode is the lowest state. The
massive bound KK modes would appear provided large $\eta$. Here we
take $\lambda=1$ for convenience. The number of bound KK modes
increases with the coupling constant $\eta$. For $\eta=0.1$, only
zero modes are bound for all $s$. For $\eta=1$, there are two bound
KK modes for $s=1$ and one bound KK mode (zero mode) for any $s\geq
3$, and the spectra of the KK modes are
\begin{eqnarray}
\begin{array}{ll}
  m_{Ln}^2 =\{0, 0.94\} \cup ~[1,\infty) & ~~~~~\text{for} ~~~~~s=1, \\
  m_{Ln}^2 =\{0\} \cup ~[1,\infty) & ~~~~~\text{for} ~~~~~s\geq3.
\end{array}
\label{spectraMSLn1CaseII}
\end{eqnarray}
For $\eta=6$, there are many bound KK modes for $s=1$ and four bound
KK modes for any $s\geq 3$, and the spectra of the KK modes are
\begin{eqnarray}
 m_{Ln}^2 &=&\{0, 10.59, 18.57, 24.37, 28.44, 31.20, 33.01,
                 \nonumber \\
       && ~ 34.17,34.89, 35.33, 35.60, 35.76, \cdots\}\nonumber\\
       &&  \cup ~[36,\infty) ~~~ \text{for} ~~s=1, \label{spectraMSLn2CaseII}\\
 m_{Ln}^2 &=&\{0, 11.89, 23.12, 32.21\} \cup ~[36,\infty)
               ~~~ \text{for} ~~s=3, \nonumber\\
 m_{Ln}^2 &=&\{0, 11.99, 23.85, 34.49\} \cup ~[36,\infty)
               ~~~ \text{for} ~~s\rightarrow\infty.~\nonumber
\end{eqnarray}
The spectra are plotted in Fig. \ref{fig_fSL_Spectra_CaseII}.  The
continuous spectrum starts at $m^2 = \eta^2$ and the KK modes
asymptotically turn into plane waves when far away from the brane,
which represent delocalized massive KK modes of fermions. It can
be seen that the spectrum structure for the single brane scenario
with $s=1$ is very different from that of the double brane
scenario with $s\geq 3$. The single brane could trap more massive
KK modes than the double brane. Noting that
\begin{eqnarray}
 V^S_L(z)=-\eta\lambda+ \eta^2 (\lambda z)^2
  +\frac{2s+1}{2s}\eta\lambda (\lambda z)^{2s}
  + {\cal{O}} ((\lambda z)^{2s+2}),\nonumber
\end{eqnarray}
we have the following simple potential for double thin brane
scenario ($s\rightarrow \infty$):
\begin{eqnarray}
 V^S_L(z)=\left\{
 \begin{array}{cc}
    -\eta\lambda+ \eta^2 (\lambda z)^2, & ~~~|\lambda z|<1 \\
    \eta^2, & ~~~|\lambda z|>1
 \end{array}\right.
\end{eqnarray}
which could be called ``the harmonic oscillator potential well with
finite depth". Similar to the square potential well with finite
depth, the spectrum can be solved and there are finite number of
bound KK modes.

\begin{figure*}[htb]
\begin{center}
\includegraphics[width=6cm,height=4cm]
{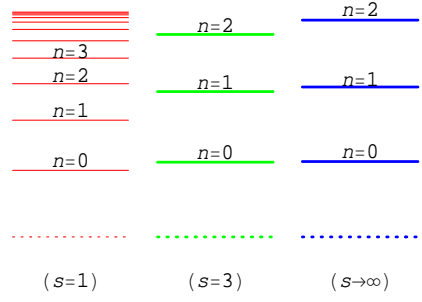}
\end{center}
\vskip -4mm \caption{(Color online) The spectra $m_{Rn}^2$ of bound
KK modes of right chiral fermions for the symmetric thick branes for
the case $F(\phi)=\tan^{1/s}(\phi/\phi_0)$ with $s=1,3$
 and $s\rightarrow\infty$.
 The parameters are set to $\lambda=1$ and $\eta=6$.
 The dashing lines denote the absent of the zero modes.}
 \label{fig_fSR_Spectra_CaseII}
\end{figure*}

The shape of the symmetric potential $V^S_{R}(z)$ (\ref{VSR_CaseII})
for right chiral fermions is more complex than that of left chiral
fermions. It has a positive value $\eta\lambda$ at $z=0$ and trends
to $\eta^2$ at $z=\pm\infty$. The shapes of the potential for
various parameters are plotted in Fig. \ref{fig_VSLR_CaseII}. For
small $\eta$, the potential for any $s$ has no well to trap bound KK
modes. With the increase of $\eta$, the potential for $s\geq 3$ will
appear a single well, while the potential for $s=1$ will first
appear a double well and then become a single well. For large
$\eta$, they are similar to a PT potential only for small $s$ (see
Fig. \ref{fig_VSR_CaseIIc}). The massless KK mode for right chiral
fermions is absent. The number of bound KK modes increases with the
coupling constant $\eta$. For $\eta=0.1$, there is no any bound KK
mode for all $s$. For $\eta=1$, there are one bound KK mode for
$s=1$ and no bound KK mode for any $s\geq 3$, and the spectra of the
KK modes are
\begin{eqnarray}
\begin{array}{ll}
  m_{Rn}^2 =\{0.94\} \cup ~[1,\infty) & ~~~~~\text{for} ~~~~~s=1, \\
  m_{Rn}^2 =\{~~\} \cup ~[1,\infty)   & ~~~~~\text{for} ~~~~~s\geq3.
\end{array}
\label{spectraMSRn1CaseII}
\end{eqnarray}
For $\eta=6$, there are many bound KK modes for $s=1$ and three
bound KK modes for any $s\geq 3$:
\begin{eqnarray}
 m_{Rn}^2 &=&\{10.59, 18.57, 24.37, 28.44, 31.20, 33.01,  \nonumber \\
               && ~ 34.17, 34.89, 35.33, 35.60, 35.76, \cdots\}\nonumber \\
          &&    \cup ~[36,\infty)~~~~~~ \text{for} ~~~~~s=1, \label{spectraMSRn2CaseII}\\
 m_{Rn}^2 &=&\{11.89, 23.12, 32.21\} \cup ~[36,\infty)
               ~~ \text{for} ~~s=3,  \nonumber\\
 m_{Rn}^2 &=&\{11.99, 23.85, 34.49\} \cup ~[36,\infty)
               ~~ \text{for} ~~s\rightarrow\infty,~~\nonumber
\end{eqnarray}
The spectra are plotted in Fig. \ref{fig_fSR_Spectra_CaseII}. The
continuous spectrum starts also at $m^2 = \eta^2$. It can be seen
that the single brane could also trap more massive KK modes than
the double brane. By comparing the mass spectra of right chiral
fermions (\ref{spectraMSRn1CaseII}) and (\ref{spectraMSRn2CaseII})
with the ones of left chiral fermions (\ref{spectraMSLn1CaseII})
and (\ref{spectraMSLn2CaseII}) for $\eta=1$ and $\eta=6$,
respectively, we come to the conclusion that the number of bound
states of right chiral fermions $N_R$ is one less than that of
left ones $N_L$, i.e., $N_R=N_L-1$. The mass spectra are almost
the same, and the only one difference is the absent of the zero
mode of right chiral fermions. Although a potential well around
the brane location appears for $s\geq3$ in Fig.
\ref{fig_VSR_CaseIIb}, we do not find any resonance.

~\\
\noindent\textbf{The asymmetric potential}\\

\begin{figure*}[htb]
\vskip -2mm
\begin{center}
\subfigure[$\eta=0.1,a=0.4$]{\label{fig_VAL_CaseIIa}
\includegraphics[width=6cm,height=4cm]{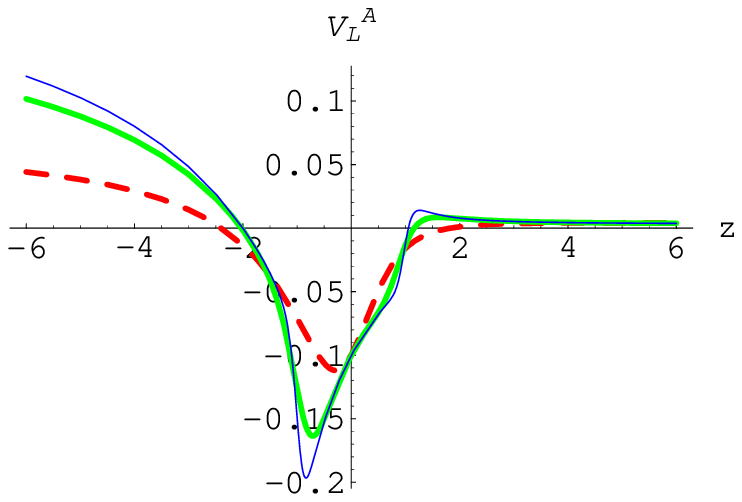}}
\subfigure[$\eta=1,a=0.02$]{\label{fig_VAL_CaseIIb}
\includegraphics[width=6cm,height=4cm]{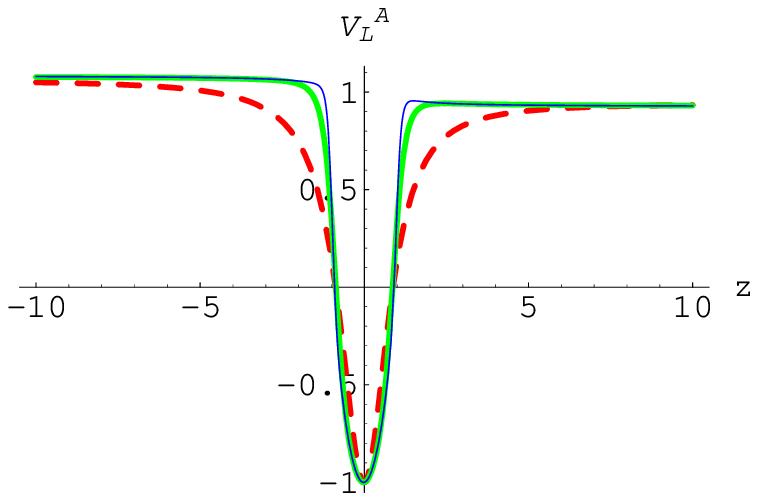}}
\subfigure[$\eta=1,a=0.2$]{\label{fig_VAL_CaseIIc}
\includegraphics[width=6cm,height=4cm]{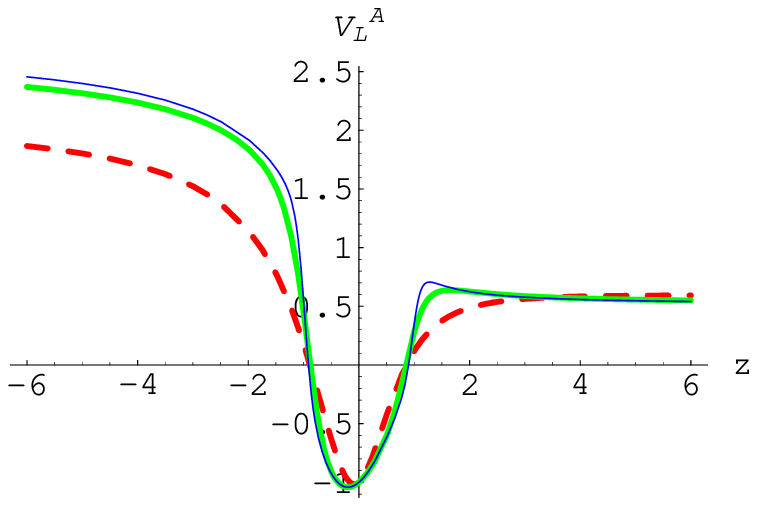}}
\subfigure[$\eta=6,a=0.2$]{\label{fig_VAL_CaseIId}
\includegraphics[width=6cm,height=4cm]{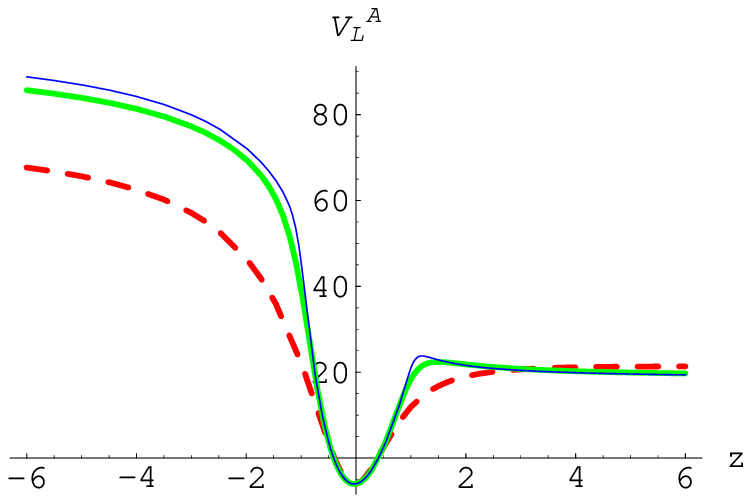}}
\end{center}
\vskip -7mm \caption{(Color online) The shapes of the asymmetric
potential $V^A_L(z)$ for the case $F(\phi)=\tan^{1/s}(\phi/\phi_0)$.
The parameter $s$ is set to $s=1$ for red dashed lines, $s=3$ for
green thick lines, and $s=7$ for blue thin lines. The parameter
$\lambda$ is set to $\lambda=1$.}
 \label{fig_VAL_CaseII}
\end{figure*}

\begin{figure*}[htb]
\vskip -2mm
\begin{center}
 \subfigure[$\eta=0.1,a=0.4$]{\label{fig_VAR_CaseIIa}
\includegraphics[width=6cm,height=4cm]{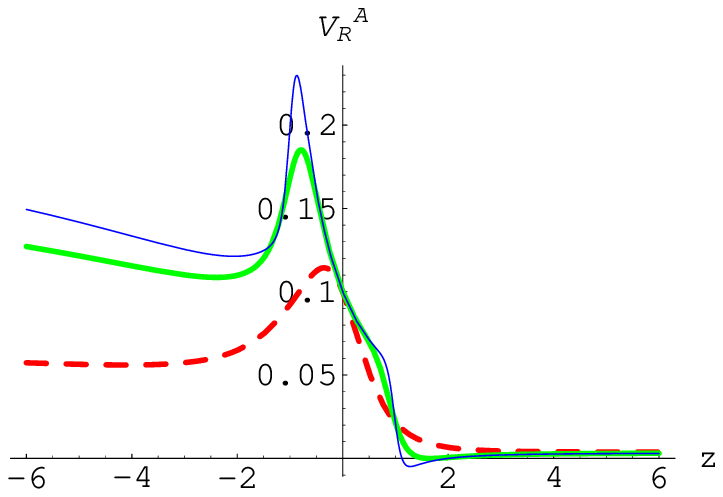}}
 \subfigure[$\eta=1,a=0.02$]{\label{fig_VAR_CaseIIb}
\includegraphics[width=6cm,height=4cm]{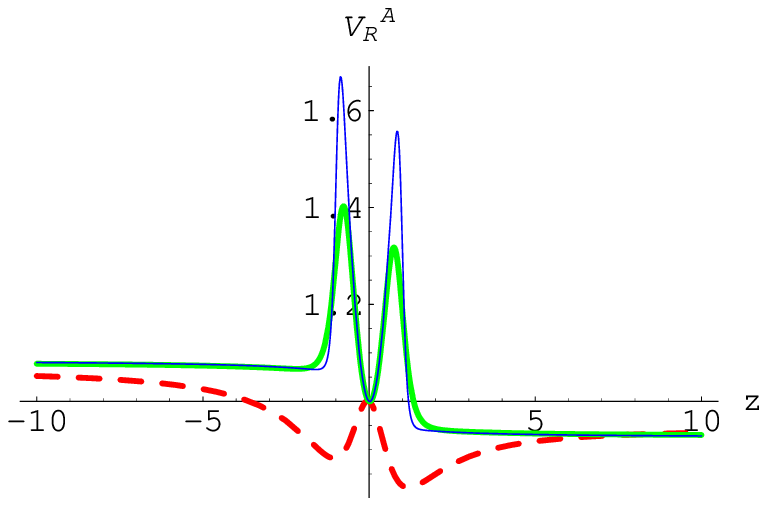}}
 \subfigure[$\eta=1,a=0.2$]{\label{fig_VAR_CaseIIc}
\includegraphics[width=6cm,height=4cm]{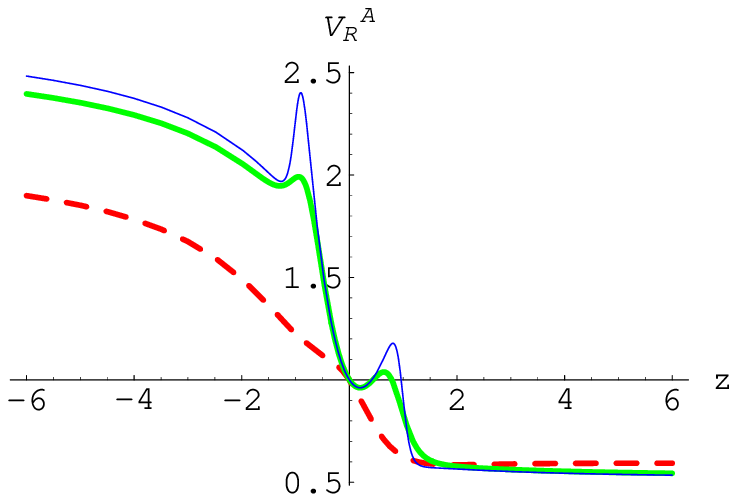}}
 \subfigure[$\eta=6,a=0.2$]{\label{fig_VAR_CaseIId}
\includegraphics[width=6cm,height=4cm]{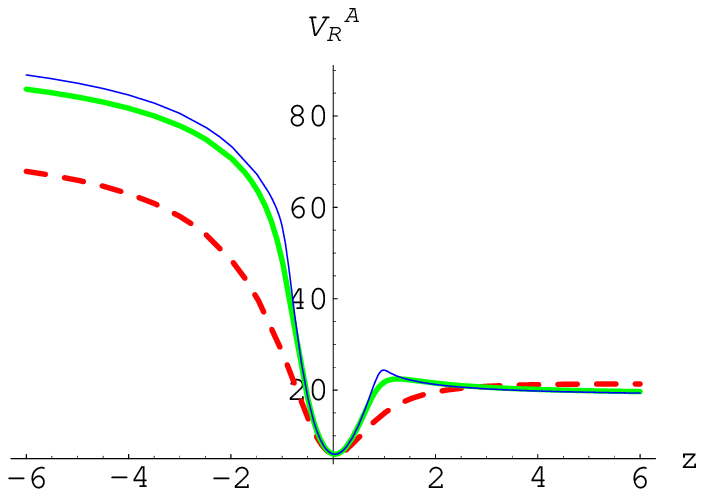}}
\end{center}
\vskip -7mm \caption{(Color online) The shapes of the asymmetric
potential $V^A_R(z)$ for the case $F(\phi)=\tan^{1/s}(\phi/\phi_0)$.
The parameters are the same as Fig. 11.}
 \label{fig_VAR_CaseII}
\end{figure*}

Next, let us turn to the potentials (\ref{VAL_CaseII}) and
(\ref{VAR_CaseII}) for the asymmetric brane. We note that, because
of the appearance of the asymmetric factor $a$, the potentials have
different limits at $z=\pm\infty$:
\begin{eqnarray}
 V^{A}_{L,R}(+\infty) = \frac{\eta^2}
  {\left(1+ a \frac{\Gamma(1/2s) \Gamma(1+1/2s)}{\lambda\;\Gamma(1/s)}
   \right)^2 }, \\
  V^{A}_{L,R}(-\infty) = \frac{\eta^2}
  {\left(1- a \frac{\Gamma(1/2s) \Gamma(1+1/2s)}{\lambda\;\Gamma(1/s)}
   \right)^2 }.
  \label{VAinftyCaseII}
\end{eqnarray}
For $s=1$ and $s\rightarrow\infty$, we have
\begin{equation}
 V^{A}_{L,R}(\pm\infty) = \frac{\eta^2}
  {\left(1\pm \frac{\pi}{2}\frac{a}{\lambda}\right)^2 }
  \label{VAinftyCaseIIs1}
\end{equation}
and
\begin{equation}
 V^{A}_{L,R}(\pm\infty) \rightarrow \frac{\eta^2}
  {\left(1\pm 2\frac{a}{\lambda}\right)^2 },
  \label{VAinftyCaseIIsinfty}
\end{equation}
respectively. The constrain condition (\ref{constraintOna}), i.e.,
$0< a \frac{\Gamma(1/2s) \Gamma(1+1/2s)}{\lambda\;\Gamma(1/s)} <1$,
implies
\begin{equation}
 0<\frac{\eta^2}{4}<V^{A}_{L,R}(\infty) <\eta^2<V^{A}_{L,R}(-\infty)<\infty.
  \label{VAinftyCaseII2}
\end{equation}
Hence, comparing with the symmetric potentials $V^{S}_{L,R}(z)$
(\ref{VSL_CaseII}) and (\ref{VSR_CaseII}), the value of the
asymmetric potentials $V^{A}_{L,R}(z)$ (\ref{VAL_CaseII}) and
(\ref{VAR_CaseII}) are enlarged at $z\rightarrow +\infty$ and
diminished at $z\rightarrow -\infty$, which would reduce the
number of the bound KK modes of left and right fermions. The
shapes of the potentials for various parameters are plotted in
Figs. \ref{fig_VAL_CaseII} and \ref{fig_VAR_CaseII}.

\begin{figure*}[htb]
\begin{center}
\includegraphics[width=6cm,height=4cm]{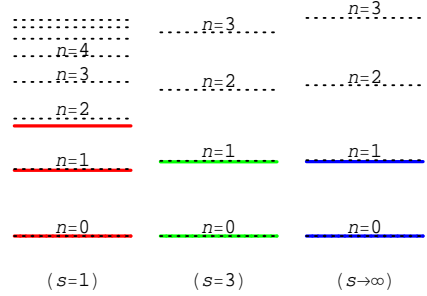}
\end{center}
\vskip -4mm \caption{(Color online) The spectra $m_{Ln}^2$ (solid
lines) of bound KK modes of left chiral fermions for the asymmetric
thick branes for the case $F(\phi)=\tan^{1/s}(\phi/\phi_0)$.
 The parameters are set to $a=0.2$, $\lambda=1$, $\eta=6$, $s=1,3$
 and $s\rightarrow\infty$. The dashing lines denote the corresponding
 spectra with $a=0$.}
 \label{fig_fAL_Spectra_CaseII}
\end{figure*}

The massless KK mode of left chiral fermions
\begin{equation}
 f_{L0}(z)    \propto
 \exp\left(\int^z d\bar{z}
     \frac{-\eta\lambda \bar{z}\;{[1+({\lambda}\bar{z})^{2s}]^{-\frac{1}{2s}}}}
       {1+a \bar{z} {~_2F_1}\left(\frac{1}{2 s},\frac{1}{s},
         1+\frac{1}{2s},-({\lambda}\bar{z})^{2s}\right)}
 \right)
  \label{zeroModefAL0CaseII}
\end{equation}
is also normalizable, and can be localized on the brane without
additional conditions. The effect of the asymmetric factor $a$ and
the coupling constant $\eta$ to the zero mode is shown in Fig.
\ref{fig_fL0_CaseII_a}. It can be seen that, with increase of
$\eta$, the effect of the asymmetric factor $a$ can be neglected.
However, the effect of $a$ to the number of bound KK modes is
remarkable. We also take $\lambda=1$ here. For $\eta=0.1$, only zero
modes of left chiral fermions are bound for all $s$ and $a$. For
$\eta=1$ and very small $a$, there is one bound massive KK mode of
both the right and left chiral fermions for $s=1$, and no bound
massive KK mode for $s\geq3$. For $\eta=1$ and large $a$, only
massless modes of left chiral fermions are bound for all $s$. For
$\eta=6$ and $a=0.2$, there are only three and two bound KK modes of
left chiral fermions for $s=1$ and $s\geq 3$, respectively, and the
spectra of the KK modes are
\begin{eqnarray}
 m_{Ln}^2 &=&\{0, 10.40, 17.42\} \cup [20.85,\infty)
               ~ \text{for} ~~s=1, \nonumber\\
 m_{Ln}^2 &=&\{0, 11.74\} \cup ~[18.75,\infty)
               ~~~~~~~~ \text{for} ~~s=3,\label{spectraMALn2CaseII}\\
 m_{Ln}^2 &=&\{0, 11.77\} \cup ~[18.37,\infty)
               ~~~~~~~~ \text{for} ~~s\rightarrow\infty,\nonumber
\end{eqnarray}
and
\begin{eqnarray}
 m_{Rn}^2 &=&\{ 10.40, 17.42\} \cup [20.85,\infty)
               ~ \text{for} ~~s=1, \nonumber\\
 m_{Rn}^2 &=&\{ 11.74\} \cup ~[18.75,\infty)
               ~~~~~~~~ \text{for} ~~s=3,\label{spectraMARn2CaseII}\\
 m_{Rn}^2 &=&\{ 11.77\} \cup ~[18.37,\infty)
               ~~~~~~~~ \text{for} ~~s\rightarrow\infty,\nonumber
\end{eqnarray}
for left and right chiral fermions, respectively. The spectra for
left chiral fermions are shown in Fig.
\ref{fig_fAL_Spectra_CaseII}. The continuous spectrum starts at
different values for different $s$. It can be seen that the
spectrum structure for the single brane case with $s=1$ is
dramatically changed by the asymmetric factor $a$, i.e., the
number of bound KK modes quickly decreases with the increase of
$a$.

\subsection{Case III: $F(\phi)=\tan^{k/s}(\phi/\phi_0)$}
\label{sec3.3}

For the case $F(\phi)=\tan^{k/s}(\phi/\phi_0)$, considering the
expression of $\phi$ (\ref{phi3}), we have $F(\phi(z))=(\lambda
z)^k$. The potential (\ref{VL}) for the asymmetric brane solution
reads as
\begin{eqnarray}
 V^A_L(z) &=& \left\{\frac{\eta^2(\lambda z)^{2k}}
                   {[1+({\lambda}z)^{2s}]^{{1}/{s}} }
           +  \frac{\eta\; a(\lambda z)^{k}}
                   {[1+({\lambda}z)^{2s}]^{{3}/{2s}} }
              \right\}\frac{1}{{\cal F}^{2}(z)}
              \nonumber \\&&
           -\frac{\eta\lambda(\lambda z)^{k-1}[k+(k-1)(\lambda z)^{2s}]}
                   {[1+({\lambda}z)^{2s}]^{1+{1}/{2s}} }
             \frac{1}{{\cal F}(z)}.\nonumber\\
   \label{VAL_CaseIII}
\end{eqnarray}
The special value $k=1$ belongs to case II considered in sub
section \ref{sec3.2}. Taking $a=0$, we will get the potential of
left chiral fermions for the symmetric brane solution:
\begin{eqnarray}
 V^S_{L,R}(z) = \frac{\eta^2(\lambda z)^{2k}}
                   {[1+({\lambda}z)^{2s}]^{{1}/{s}} }
             \mp\frac{\eta\lambda(\lambda z)^{k-1}[k+(k-1)(\lambda z)^{2s}]}
                   {[1+({\lambda}z)^{2s}]^{1+/{1}/{2s}} }.
   \label{VSL_CaseIII}
\end{eqnarray}

~\\
\noindent\textbf{The symmetric potential}\\

\begin{figure*}[htb]
\vskip -5mm
\begin{center}
\subfigure[$\eta=\lambda=1,k=-1$]{\label{fig_VSL_CaseIII1a}
\includegraphics[width=6cm,height=4cm]{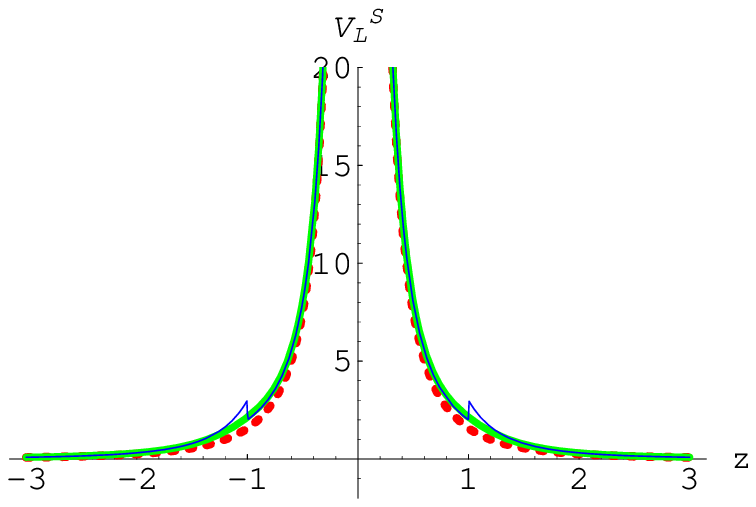}}
\subfigure[$\eta=\lambda=1,k=-1$]{\label{fig_VSR_CaseIII1a}
\includegraphics[width=6cm,height=4cm]{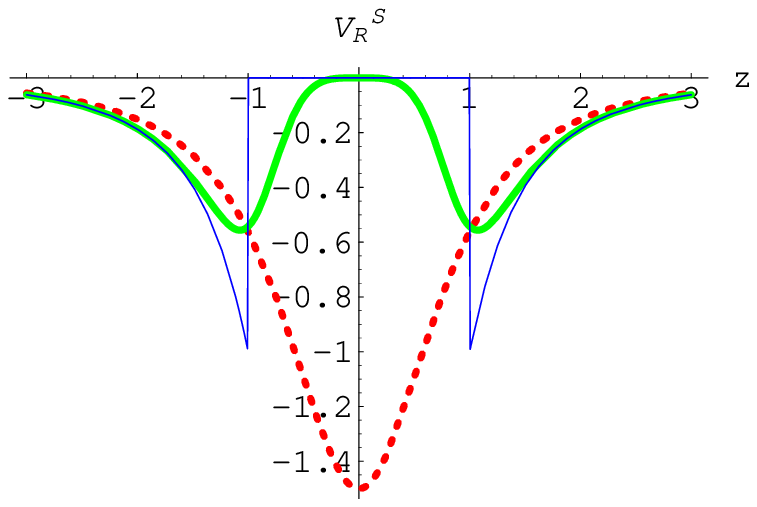}}
\subfigure[$\eta=1,\lambda=1.1,k=-1$]{\label{fig_VSL_CaseIII1b}
\includegraphics[width=6cm,height=4cm]{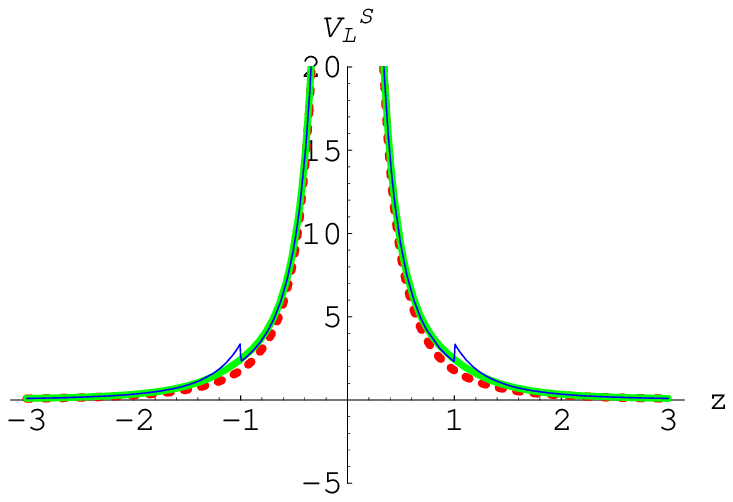}}
\subfigure[$\eta=1,\lambda=1.1,k=-1$]{\label{fig_VSR_CaseIII1b}
\includegraphics[width=6cm,height=4cm]{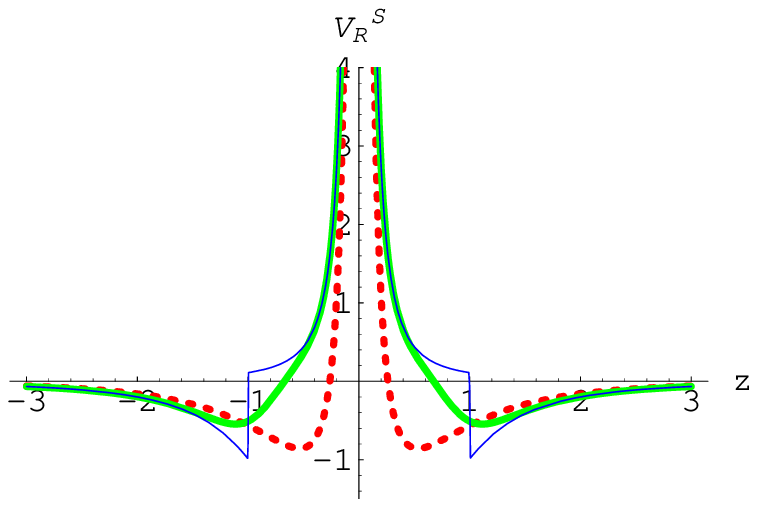}}
\subfigure[$\eta=\lambda=1,k=-3$]{\label{fig_VSL_CaseIII1c}
\includegraphics[width=6cm,height=4cm]{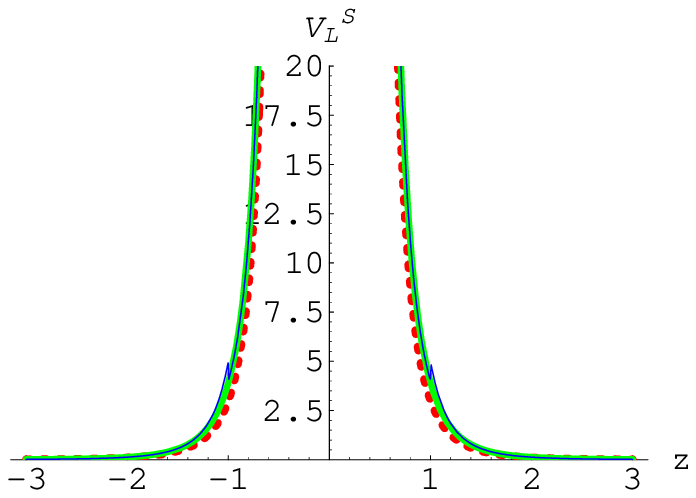}}
\subfigure[$\eta=\lambda=1,k=-3$]{\label{fig_VSR_CaseIII1c}
\includegraphics[width=6cm,height=4cm]{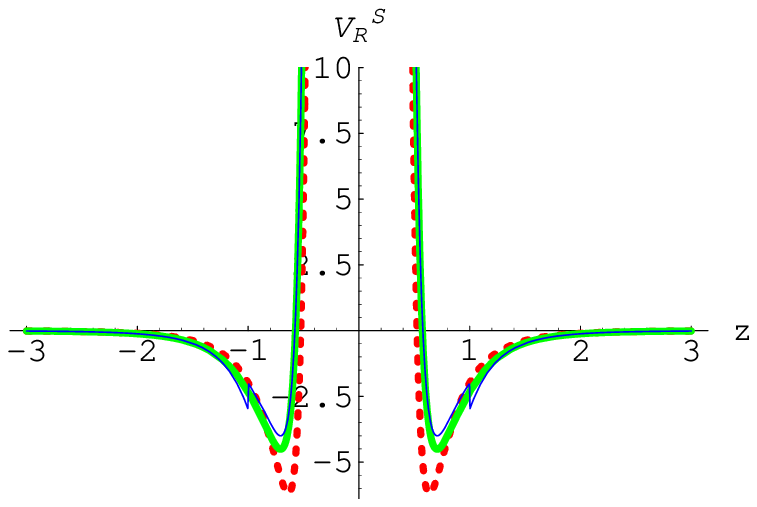}}
\end{center}
\vskip -7mm \caption{(Color online) The shapes of the symmetric
potentials $V^S_L(z)$ and $V^S_R(z)$ for the case
$F(\phi)=\tan^{k/s}(\phi/\phi_0)$ with odd $k<0$. The parameter $s$
is set to $s=1$ for red dashed lines, $s=3$ for green thick lines,
and $s\rightarrow\infty$ for blue thin lines.}
 \label{fig_VSLR_CaseIII1}
\end{figure*}

\begin{figure*}[htb]
\vskip -5mm
\begin{center}
\subfigure[$k=3$]{\label{fig_VSL_CaseIII2a}
\includegraphics[width=6cm,height=4cm]{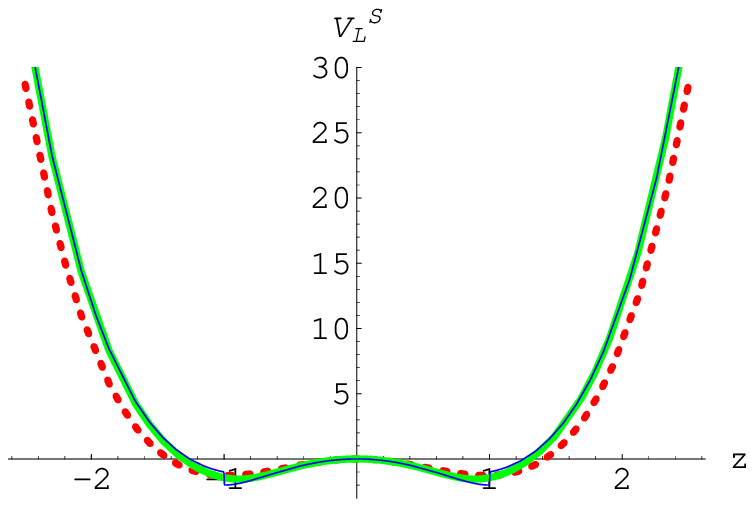}}
\subfigure[$k=3$]{\label{fig_VSR_CaseIII2a}
\includegraphics[width=6cm,height=4cm]{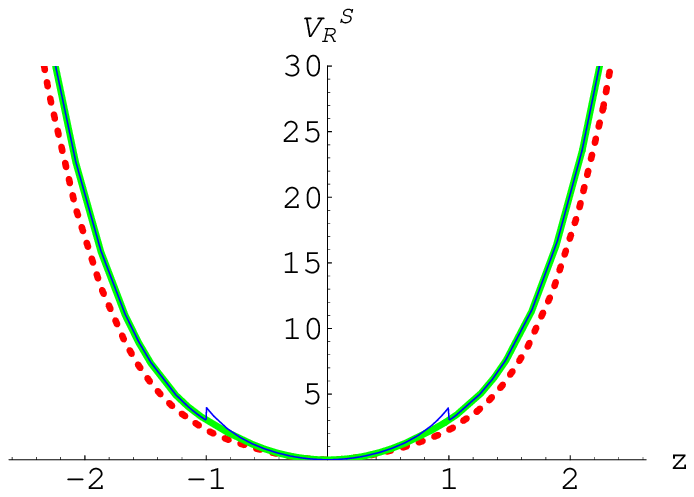}}
\subfigure[$k=11$]{\label{fig_VSL_CaseIII2c}
\includegraphics[width=6cm,height=4cm]{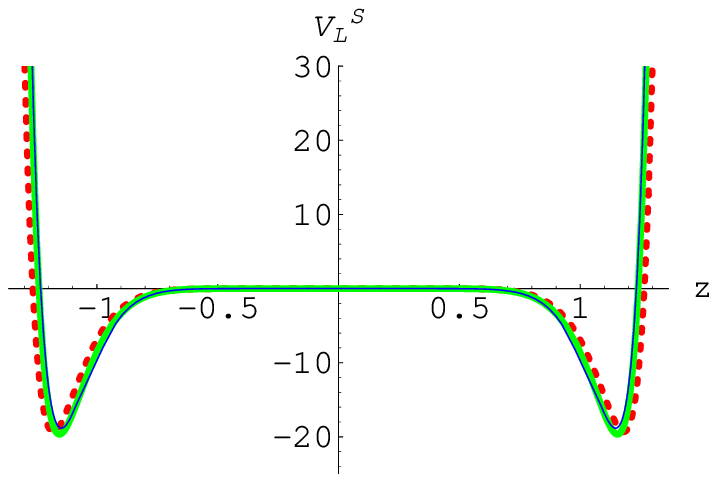}}
\subfigure[$k=11$]{\label{fig_VSR_CaseIII2b}
\includegraphics[width=6cm,height=4cm]{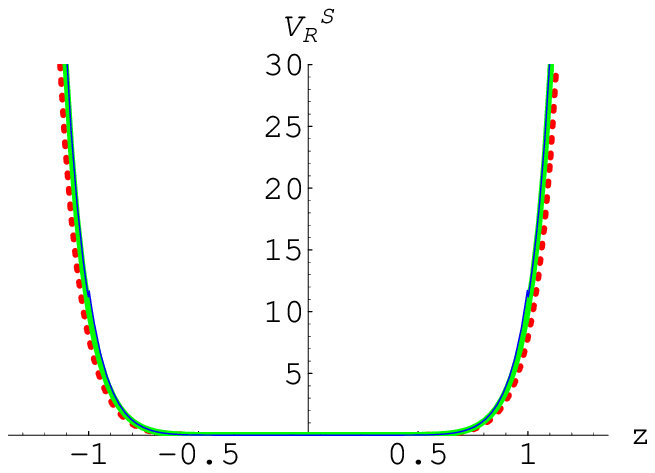}}
\subfigure[$k\rightarrow\infty$]{\label{fig_VSL_CaseIII2c}
\includegraphics[width=6cm,height=4cm]{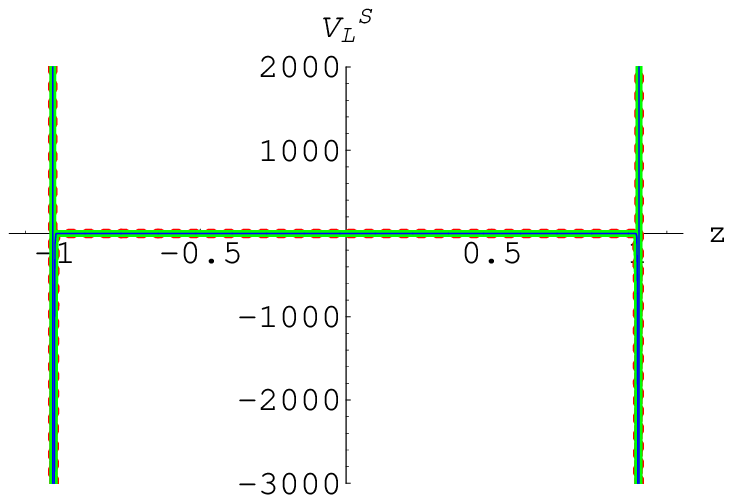}}
\subfigure[$k\rightarrow\infty$]{\label{fig_VSR_CaseIII2c}
\includegraphics[width=6cm,height=4cm]{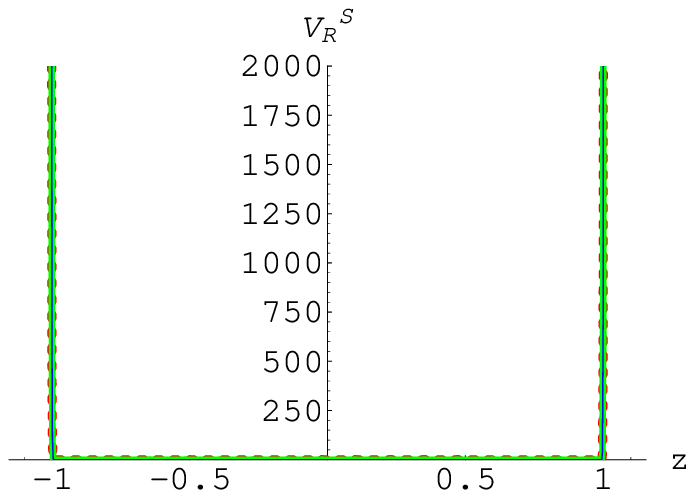}}
\end{center}
\vskip -7mm \caption{(Color online) The shapes of the symmetric
potentials $V^S_L(z)$ and $V^S_R(z)$ for the case
$F(\phi)=\tan^{k/s}(\phi/\phi_0)$ with odd $k\geq3$. The parameter
$s$ is set to $s=1$ for red dashed lines, $s=3$ for green thick
lines, and $s\rightarrow\infty$ for blue thin lines. The other
parameters are set to $\eta=\lambda=1$ and $k=3,11,\infty$.}
 \label{fig_VSLR_CaseIII2}
\end{figure*}

Let us first analyze the asymptotic property of the symmetric
potential. When $z\rightarrow\infty$, $1 + (\lambda z)^{2
s}\rightarrow(\lambda z)^{2 s}$, we have
\begin{eqnarray}
 V^S_{L,R}(z\rightarrow\infty) &\rightarrow& \eta^2 (\lambda z)^{2k-2}
          \mp \eta \lambda k (\lambda z)^{k-2-2s}\nonumber\\
       &&   \mp \eta \lambda (k-1) (\lambda z)^{k-2} \nonumber\\
  &\rightarrow&\left\{
  \begin{array}{ll}
    \eta^2 (\lambda z)^{2k-2} \rightarrow\infty
               & ~~ \text{for}~~ k>1 \\
    \eta^2 >0  & ~~ \text{for}~~ k=1 \\
    0 & ~~ \text{for}~~ k< 1
  \end{array}\right..~~~~~
\end{eqnarray}
When $z\rightarrow0$, $1 + (\lambda z)^{2 s}\rightarrow1$, we have
\begin{eqnarray}
 &&V^S_{L,R}(z\rightarrow0) \rightarrow
   \eta^2 (\lambda z)^{2k} \left(1-\frac{1}{s}({\lambda}z)^{2s}\right)
   \nonumber \\
   &&\mp\eta\lambda(\lambda z)^{k-1}[k+(k-1)(\lambda z)^{2s}]
   \left(1-\frac{2s+1}{2s}({\lambda}z)^{2s}\right) \nonumber\\
  &&\rightarrow\left\{
  \begin{array}{ll}
    0                 & ~~ \text{for}~~ k>1 \\
    \mp \eta \lambda  & ~~ \text{for}~~ k=1 \\
    \frac{\eta(\eta\pm\lambda)}{(\lambda z)^{2}}
       + (-\eta^2 \pm \frac{1}{2}\eta\lambda)\delta_{s,1}
                      & ~~ \text{for}~~ k=-1\\
    \infty            & ~~ \text{for}~~ k<-1 \\
  \end{array}\right. .
\end{eqnarray}
For $\eta>0$, $\lambda>0$ and $k=-1$,
\begin{eqnarray}
 V^S_{L}(z\rightarrow0) &\rightarrow&
      \infty  \\
 V^S_{R}(z\rightarrow0) &\rightarrow&  \left\{
  \begin{array}{cl}
    \infty                 & ~~~~ \text{for}~~~~ \eta\neq\lambda \\
    -\frac{3}{2}\eta^2     & ~~~~ \text{for}~~~~ \eta=\lambda, s=1 \\
    0                      & ~~~~ \text{for}~~~~ \eta=\lambda, s>1
  \end{array}\right. .
  \end{eqnarray}
The shapes of the symmetric potentials $V^S_L(z)$ and $V^S_R(z)$ for
the case $F(\phi)=\tan^{k/s}(\phi/\phi_0)$ with odd $k<0$ and $k>1$
are plotted in Figs. \ref{fig_VSLR_CaseIII1} and
\ref{fig_VSLR_CaseIII2}, respectively. For the case $k<0$, the
symmetric potential $V^S_L(z)$ of left chiral fermions has no well
and can not trap any bound KK modes. For right chiral fermions, the
symmetric potential $V^S_R(z)$ with $k<-1$ or $k=-1$,
$\eta\neq\lambda$, has a double well and a infinite high bar, which
can also not trap the massless mode. However, the case $k=-1$,
$\eta=\lambda>0$ is very special, for which $V^S_R(z)$ has a single
well (for $s=1$) or a double well (for $s\geq3$) (see Fig.
\ref{fig_VSR_CaseIII1a}). The bound KK modes corresponding to the
potentials shown in Fig. \ref{fig_VSR_CaseIII1a} for $s=1$, $s=3$
and $s\rightarrow\infty$ have mass square $-0.636$, $-0.1485$ and
$-0.119$, respectively. Because all the potentials with $k<0$ can
not trap zero modes, or even more, some of them could result in
tachyonic KK modes with $m^2<0$, we do not consider the
corresponding kink-fermion coupling of the type
$\eta\overline{\Psi}\tan^{k/s}(\phi/\phi_0)\Psi$ with $k<0$.

\begin{figure*}[htb]
\vskip -5mm
\begin{center}
\subfigure[$k=3$]{\label{fig_Mn2_CaseIII1a}
\includegraphics[width=6cm,height=4cm]
     {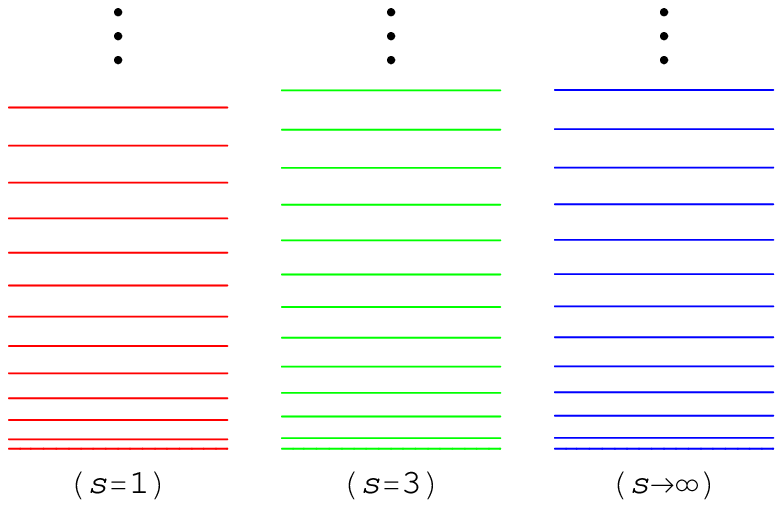}}
\hskip 5mm\subfigure[$k=11$]{\label{fig_Mn2_CaseIII1b}
\includegraphics[width=6cm,height=4cm]
 {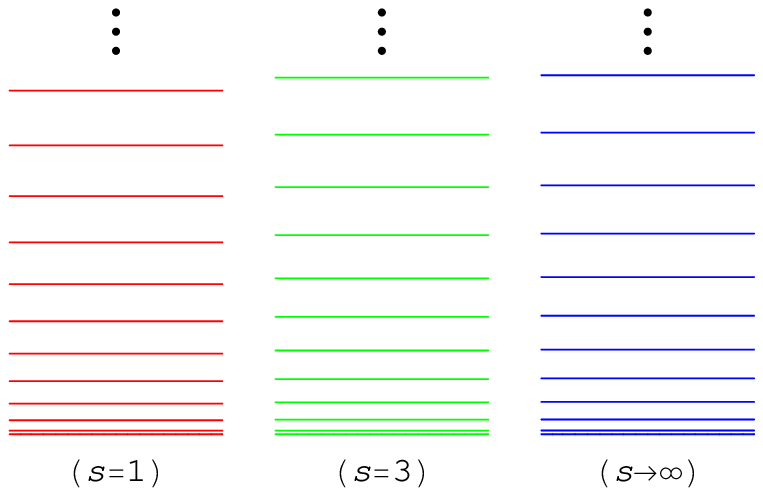}}
\end{center}
\vskip -7mm \caption{(Color online) The spectra $m_{Ln}^2$ of left
chiral fermions for the case $F(\phi)=\tan^{k/s}(\phi/\phi_0)$ with
$k=3$ and $11$. The parameters are set to $a=0$, $\eta=\lambda=1$,
$s=1$, 3, and $\infty$.}
 \label{fig_Mn2_CaseIII1}
\end{figure*}

For $k>1$, which is the case we are interesting in here, the
potentials $V^S_{L,R}(z)$ trend to infinite when far away from the
brane and vanish at $z=0$, which shows that there exist infinite
discrete bound KK modes. We note from Fig. \ref{fig_VSLR_CaseIII2}
that the influence of $s$ is not important. For examples, the
spectra for $k=3$ and $k=11$ are calculated as
\begin{eqnarray}
 m_{Ln}^2&=&\{0, 1.57, 4.80, 8.44, 12.61, 17.17, 22.08, \nonumber \\
    && 27.30, 32.78, 38.52, 44.49,
       \cdots\} ~~~ \text{for} ~s=1, \nonumber\\
 m_{Ln}^2 &=&\{0, 1.76, 5.39, 9.33, 13.73, 18.56, 23.70,  \nonumber \\
    && 29.14, 34.85, 40.80, 46.97,
        \cdots\} ~~~ \text{for} ~s=3,\\
 m_{Ln}^2 &=&\{0, 1.81, 5.52, 9.44, 13.77, 18.65, 23.81, \nonumber \\
    &&  29.19, 34.93, 40.88, 47.02,
       \cdots\}~~~ \text{for} ~s\rightarrow\infty,\nonumber
\end{eqnarray}
and
\begin{eqnarray}
 m_{Ln}^2 &=&\{0, 1.86, 7.30, 16.02, 27.68, 42.03, 58.93, \nonumber \\
    &&78.30, 100.08, 124.22, \cdots\}
               ~~~ \text{for} ~s=1, \nonumber\\
 m_{Ln}^2 &=&\{0, 1.93, 7.59, 16.66, 28.77, 43.67, 61.21, \nonumber \\
    && 81.30, 103.87, 128.88, \cdots\}
               ~~~ \text{for} ~s=3,\\
 m_{Ln}^2 &=&\{0, 1.96, 7.69, 16.87, 29.12, 44.15, 61.81, \nonumber \\
    && 82.00, 104.68, 129.82, \cdots\}
               ~~~ \text{for} ~s\rightarrow\infty,\nonumber
\end{eqnarray}
respectively, and shown in Fig. \ref{fig_Mn2_CaseIII1}.

\begin{figure*}[htb]
\vskip -5mm
\begin{center}
\subfigure[$m_{Ln}^2$]{\label{fig_Mn2_CaseIII2a}
\includegraphics[width=6cm,height=4cm]
     {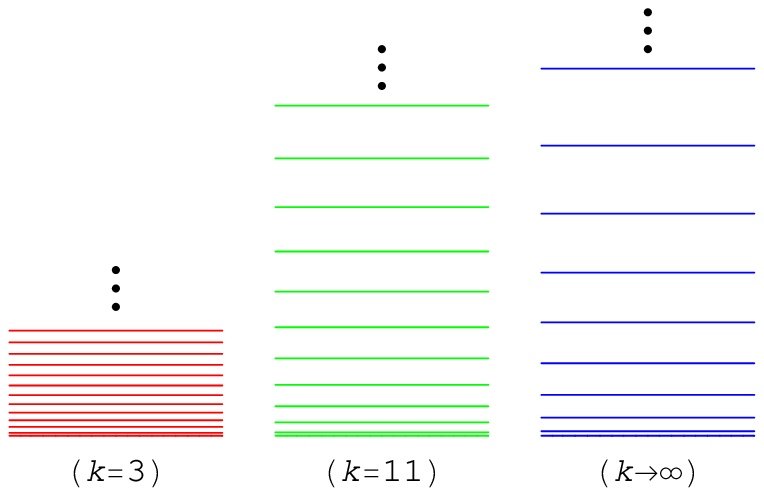}}
 \hskip 5mm
\subfigure[$m_{Ln}$]{\label{fig_Mn2_CaseIII2b}
\includegraphics[width=6cm,height=4cm]
     {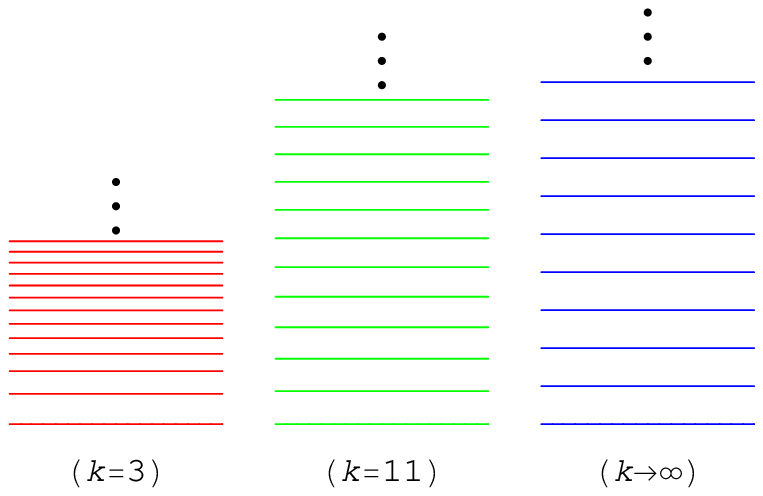}}
\end{center}
\vskip -4mm \caption{(Color online) The spectra  $m_{Ln}^2$ and
$m_{Ln}$ of left chiral fermions for the case
$F(\phi)=\tan^{k/s}(\phi/\phi_0)$ with $k=3,$ 11 and $\infty$. The
parameters are set to $a=0$, $s=\eta=\lambda=1$.}
 \label{fig_Mn2_CaseIII2}
\end{figure*}

With the increase of $k$, the potentials vanish in a more wide range
around $z=0$. Especially, for $k\rightarrow\infty$, we get an
infinite deep square well for right hand fermions:
\begin{eqnarray}
 V^S_{R} = \left\{
 \begin{array}{cc}
   \infty, & ~~~~|z|>{1}/{\lambda} \\
   0, &      ~~~~|z|<{1}/{\lambda}
 \end{array}
 \right. ,
\end{eqnarray}
The KK modes and the spectrum reads as
\begin{eqnarray}
 f_{Rn}=\left\{\begin{array}{cc}
     \sqrt{\lambda}\cos({n\pi\lambda z}/{2}), & ~~n=1,3,5,\cdots \\
     \sqrt{\lambda}\sin({n\pi\lambda z}/{2}), & ~~n=2,4,6,\cdots
        \end{array}\right.
 \label{fRn_CaseIIIa}
\end{eqnarray}
\begin{eqnarray}
 m_{Rn}^2=\left(\frac{\pi}{2}\lambda n \right)^2. ~~~(n=1,2,3,\cdots)
 \label{mn2_CaseIIIa}
\end{eqnarray}
The numeric result for $k=13111$, $s=\eta=\lambda=1$ is
\begin{eqnarray}
 m_{Rn}^2 =\{2.46, 9.85, 22.16, 39.39, 61.54,
             88.62, 
             \cdots\},
\end{eqnarray}
from which we have
\begin{eqnarray}
 \frac{m_{Rn}}{m_{R1}} =\{1, 2.001, 3.001, 4.002, 5.002,
        6.002, 
        \cdots\}.
\end{eqnarray}
For left hand fermions, the spectrum also takes the form
(\ref{mn2_CaseIIIa}) but with $n=0,1,2,\cdots$, the KK modes can be
calculated from Eqs. (\ref{CoupleEq1b}) and (\ref{fRn_CaseIIIa}) as:
\begin{eqnarray}
 f_{Ln}=\left\{\begin{array}{ll}
   -\sqrt{\lambda}\sin({n\pi\lambda z}/{2}), & ~~n=1,3,5,\cdots \\
   \left\{
          \begin{array}{cl}
                \lambda/2,  & ~|{z}|<1/\lambda \\
                0,  & ~|{z}|>1/\lambda
          \end{array}\right.,
       & ~~n=0 \\
   -\sqrt{\lambda}\cos({n\pi\lambda z}/{2}), & ~~n=2,4,6,\cdots
        \end{array}\right.
 \label{fLn_CaseIIIa}
\end{eqnarray}
The comparing of spectra of left chiral fermions for different $k$
is shown in Fig. \ref{fig_Mn2_CaseIII2}.

~\\ \noindent\textbf{Asymmetric branes}\\

\begin{figure*}[htb]
\vskip -8mm
\begin{center}
\subfigure[$k=3,s=1$]{\label{fig_VSL_CaseIII2a}
\includegraphics[width=6cm,height=4cm]{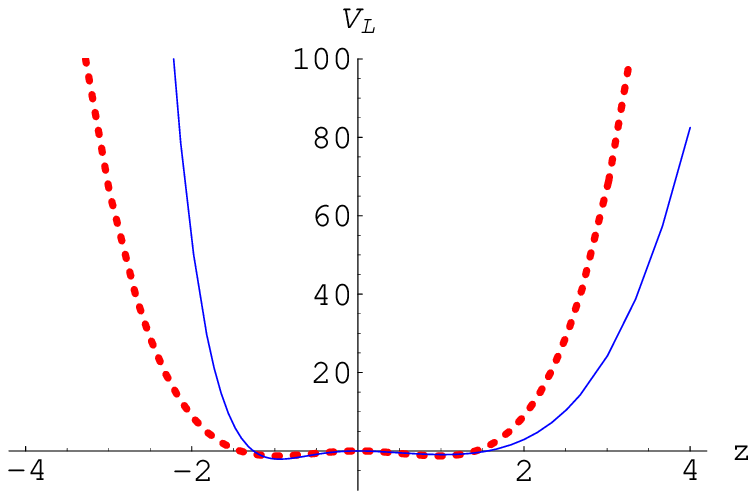}}
\subfigure[$k=3,s\rightarrow\infty$]{\label{fig_VSR_CaseIII2a}
\includegraphics[width=7cm,height=4cm]{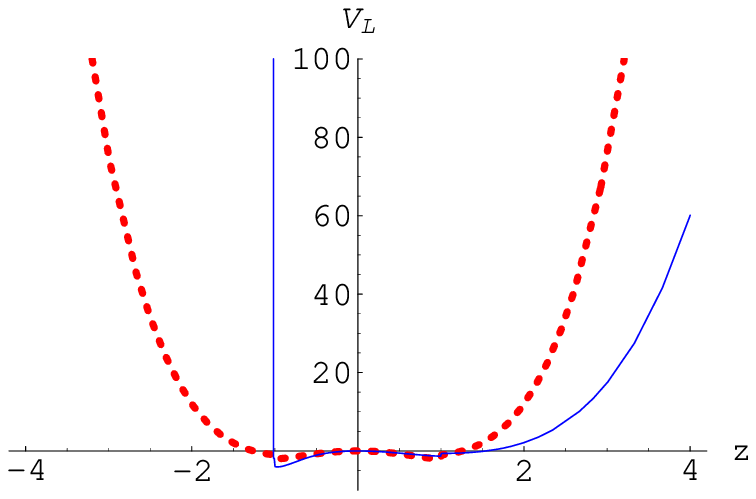}}
\subfigure[$k=11,s=1$]{\label{fig_VSL_CaseIII2b}
\includegraphics[width=6cm,height=4cm]{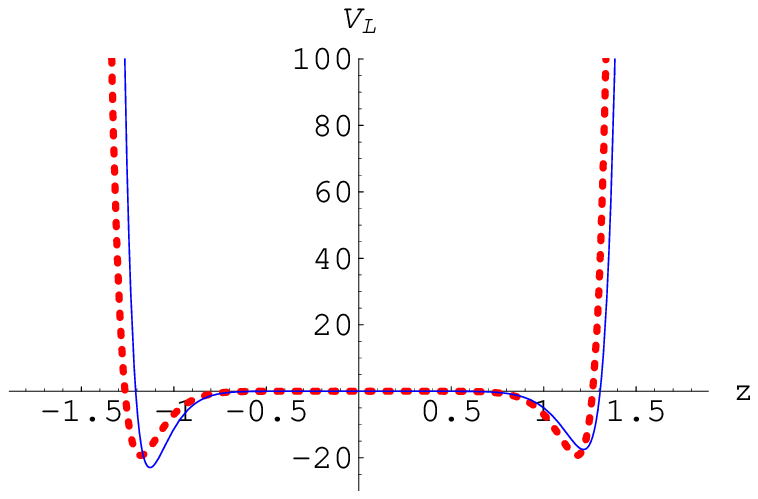}}
\subfigure[$k=11,s\rightarrow\infty$]{\label{fig_VSR_CaseIII2b}
\includegraphics[width=6cm,height=4cm]{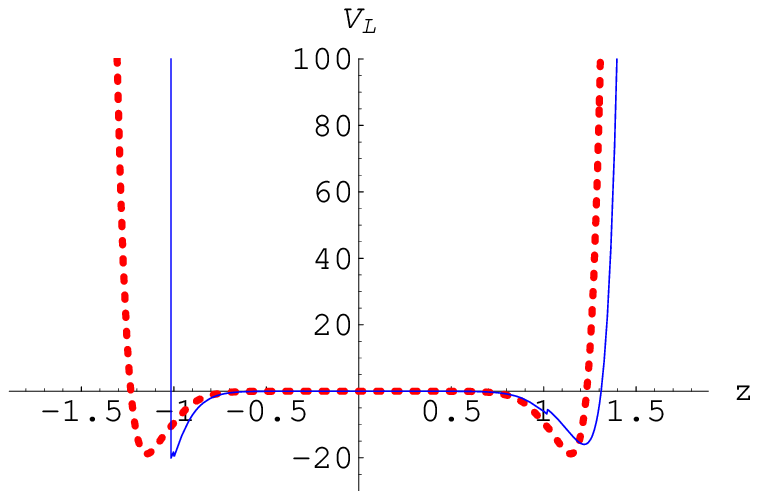}}
\subfigure[$k\rightarrow\infty,s=1$]{\label{fig_VSL_CaseIII2c}
\includegraphics[width=6cm,height=4cm]{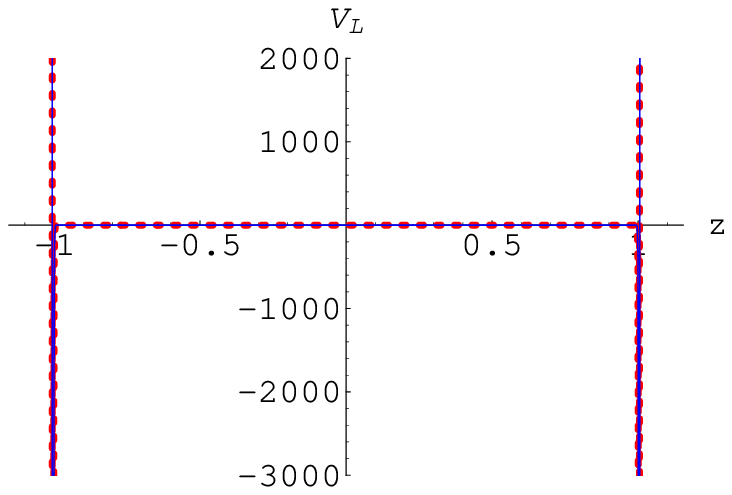}}
\subfigure[$k\rightarrow\infty,s\rightarrow\infty$]{\label{fig_VSR_CaseIII2c}
\includegraphics[width=6cm,height=4cm]{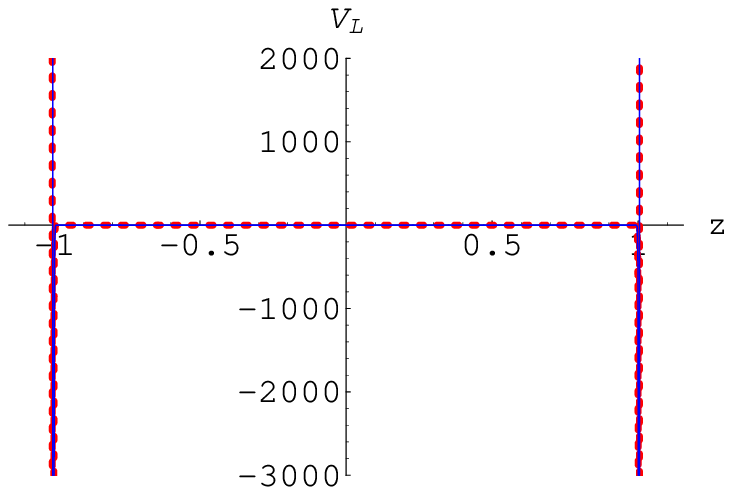}}
\end{center}
\vskip -7mm \caption{(Color online) Comparison of the asymmetric
potential $V^A_L(z)$ (blue thin lines) with the symmetric potential
$V^S_L(z)$ (red dashed lines) for the case
$F(\phi)=\tan^{k/s}(\phi/\phi_0)$ with different $k$ and $s$. The
parameters are set to $\eta=\lambda=1$, $a=0$ for red dashed lines
and $a=0.5$ for blue thin lines, $s=1$ and $\infty$, $k=3,7$
and$\infty$.}
 \label{fig_VAL_CaseIII}
\end{figure*}

\begin{figure*}[htb]
\vskip -5mm
\begin{center}
\includegraphics[width=6cm,height=4cm]
     {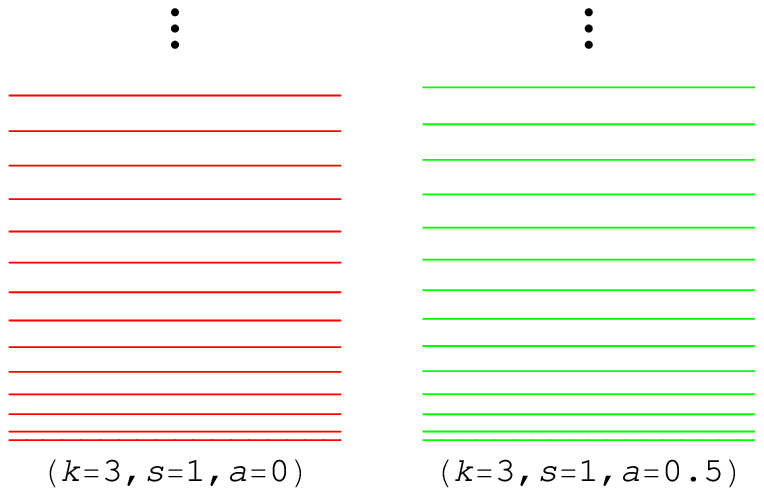}
\hskip 5mm
\includegraphics[width=6cm,height=4cm]
 {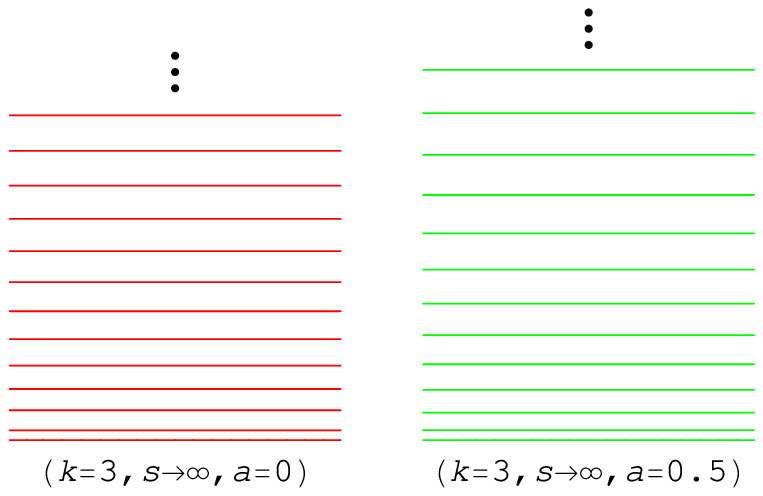}
\vskip 5mm
\includegraphics[width=6cm,height=4cm]
     {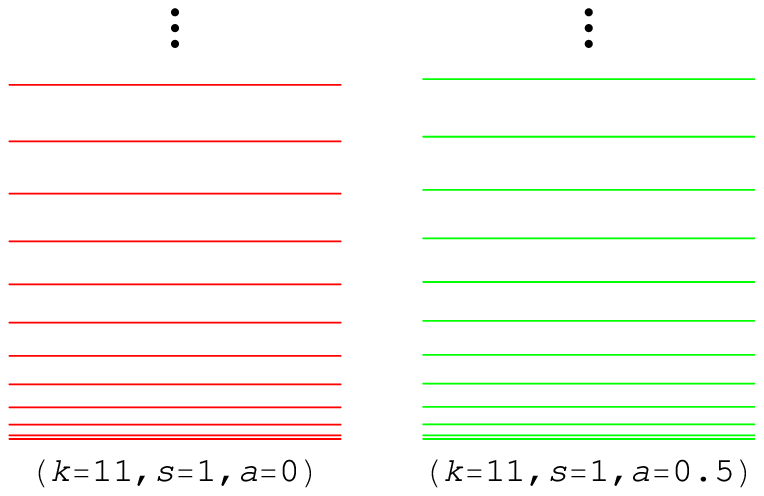}
\hskip 5mm
\includegraphics[width=6cm,height=4cm]
 {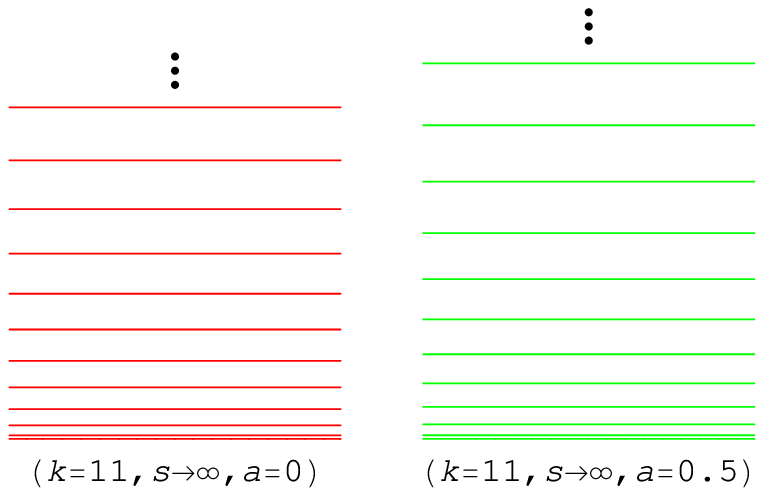}
\end{center}
\vskip -4mm \caption{(Color online) Comparison of the spectra
$m_{Ln}^2$ of the asymmetric potential $V^A_L(z)$ ($a=0.5$) with
that of the symmetric potential $V^S_L(z)$ ($a=0$) for the case
$F(\phi)=\tan^{k/s}(\phi/\phi_0)$ with $k=3,11$ and $s=1,\infty$.
The parameters are set to $\eta=\lambda=1$.}
 \label{fig_Mn2_CaseIII}
\end{figure*}

At last, we consider the asymmetric potential (\ref{VAL_CaseIII})
for left chiral fermions and the corresponding asymmetric potential
(\ref{VR}) for right chiral fermions. The asymptotic property of
them is analyzed as follows. When $z\rightarrow\pm\infty$, $1 +
(\lambda z)^{2 s}\rightarrow(\lambda z)^{2 s}$,
${\cal{F}}(z)\rightarrow \left(1\pm a \frac{\Gamma(1/2s)
\Gamma(1+1/2s)}{\lambda\;\Gamma(1/s)}\right)$, we have
\begin{eqnarray}
 V^A_{L,R}(z)
  &\rightarrow&\left\{
  \begin{array}{ll}
    \frac{\eta^2 (\lambda z)^{2k-2}}
         {\left(1\pm a \frac{\Gamma(1/2s) \Gamma(1+1/2s)}
                            {\lambda\;\Gamma(1/s)}\right)^2 }
    \rightarrow\infty  & \text{for}~ k>1 \\
    \frac{\eta^2}
         {\left(1\pm a \frac{\Gamma(1/2s) \Gamma(1+1/2s)}
    {\lambda\;\Gamma(1/s)}\right)^2 }>0  & \text{for}~ k=1 \\
    0 & \text{for}~ k< 1
  \end{array}\right. .
\end{eqnarray}
Since $V^S_{L,R}(z\rightarrow\pm\infty)$ are finite for $k\leq3$, we
only consider the case $k\geq3$ here, for which
$V^S_{L,R}(z\rightarrow0)\rightarrow0$. Comparison of the asymmetric
potential $V^A_L(z)$ with the symmetric one $V^S_L(z)$ for the case
$F(\phi)=\tan^{k/s}(\phi/\phi_0)$ with different $k$ and $s$ is
shown in Fig. \ref{fig_VAL_CaseIII}. We see that, for a fixed finite
$k$, the difference of the two potentials would become large with
increase of $s$. For $s\rightarrow\infty$, the difference is
largest. While for a fixed $s$, the difference of the two potentials
would become small with increase of $k$. The spectra for $k=3$ and
$k=11$ are calculated as
\begin{eqnarray}
 m_{Ln}^2 &=&\{0, 1.59, 4.77, 8.50, 12.7, 17.4, 22.4, 27.7,
   \nonumber\\ &&
       33.3, 39.2, 45.4, 51.7, \cdots\},
               ~ (s=1) \\
 m_{Ln}^2 &=&\{0, 1.83, 5.09, 9.27, 14.0, 19.4, 25.2, 31.5,
 \nonumber\\ &&
       38.2, 45.3, 52.7, 60.4, \cdots\},
               ~ (s\rightarrow\infty)
\end{eqnarray}
and
\begin{eqnarray}
 m_{Ln}^2 &=&\{0, 1.88, 7.40, 16.2, 28.1, 42.6, 59.8, 79.5,
 \nonumber\\ &&
       102, 126, 153, 182, \cdots\},
               ~ (s=1) \\
 m_{Ln}^2 &=&\{0, 2.09, 8.22, 18.1, 31.4, 47.8, 67.5, 90.3,
 \nonumber\\ &&
       116, 145, 177, 212, \cdots\},
               ~ (s\rightarrow\infty)
\end{eqnarray}
respectively, where $a=0.5$, and the comparing with that of the
symmetric potential is shown in Fig. \ref{fig_Mn2_CaseIII}. For
$k\rightarrow\infty$, the difference between $V^A_L(z)$ and
$V^S_L(z)$ disappears, and the spectrum is $m_{Ln}=n\lambda{\pi}
/{2} ~(n=0,1,2,3,\cdots)$.

\begin{figure*}[htb]
\vskip -5mm
\begin{center}
\subfigure[$k=3,a=0$]{\label{fig_fAL_CaseIII_eta_a}
\includegraphics[width=6cm,height=4cm]{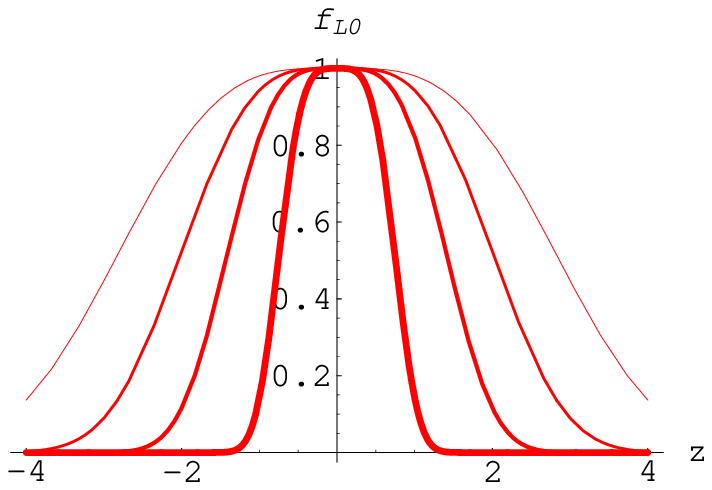}}
\subfigure[$k=11,a=0$]{\label{fig_fAL_CaseIII_eta_b}
\includegraphics[width=6cm,height=4cm]{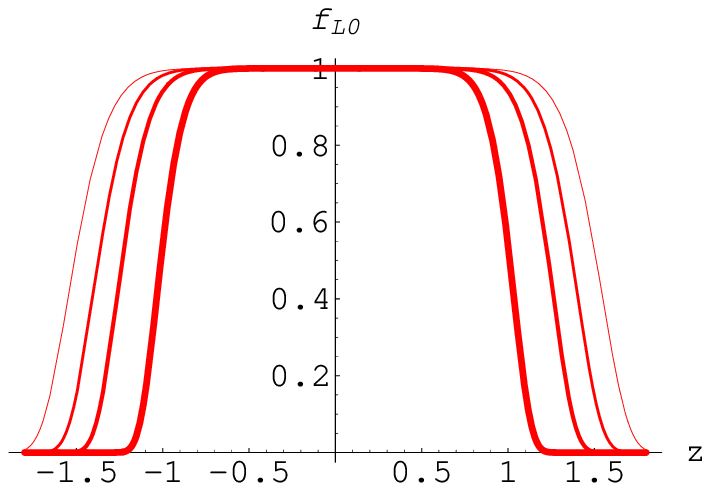}}
\subfigure[$k=3,a=0.5$]{\label{fig_fAL_CaseIII_eta_c}
\includegraphics[width=6cm,height=4cm]{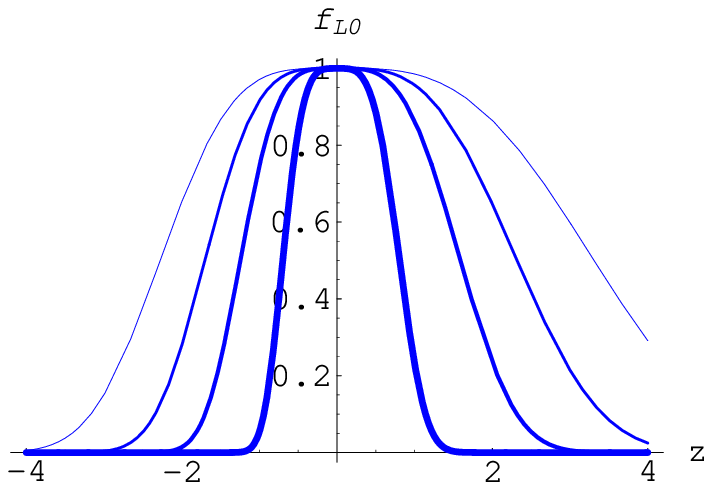}}
\subfigure[$k=11,a=0.5$]{\label{fig_fAL_CaseIII_eta_d}
\includegraphics[width=6cm,height=4cm]{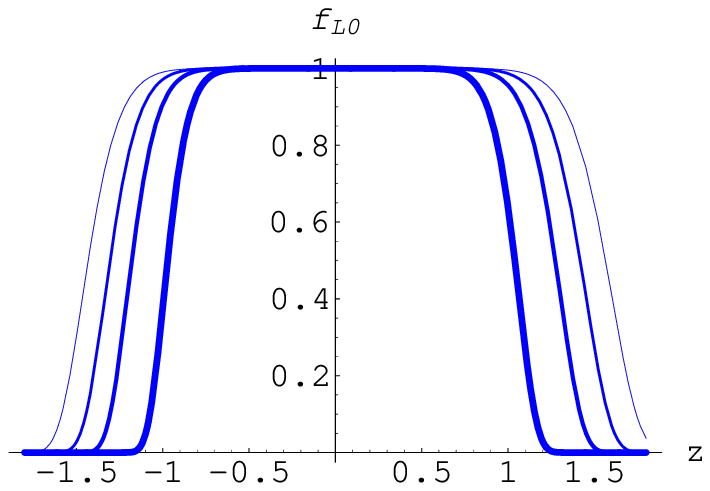}}
\end{center}
\vskip -7mm \caption{(Color online) Comparison of the zero modes of
the asymmetric potential $V^A_L(z)$ (down two) with that of the
symmetric potential $V^S_L(z)$ (up two) for the case
$F(\phi)=\tan^{k/s}(\phi/\phi_0)$ with different $k$ and $\eta$. The
parameters are set to $s=1$, $\lambda=1$, $a=0,0.5$, $k=3,11$,
$\eta=0.1,0.3,1,10$. The thickness of lines increases with $\eta$.}
 \label{fig_fAL_CaseIII_eta}
\end{figure*}

\begin{figure*}[htb]
\begin{center}
\subfigure[$a=0,s=1$]{\label{fig_fAL_CaseIIIa}
\includegraphics[width=6cm,height=4cm]{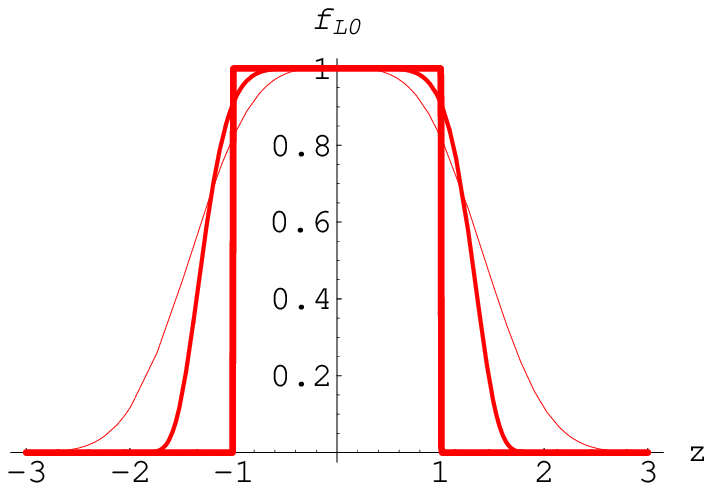}}
\subfigure[$a=0,s\rightarrow\infty$]{\label{fig_fAL_CaseIIIb}
\includegraphics[width=6cm,height=4cm]{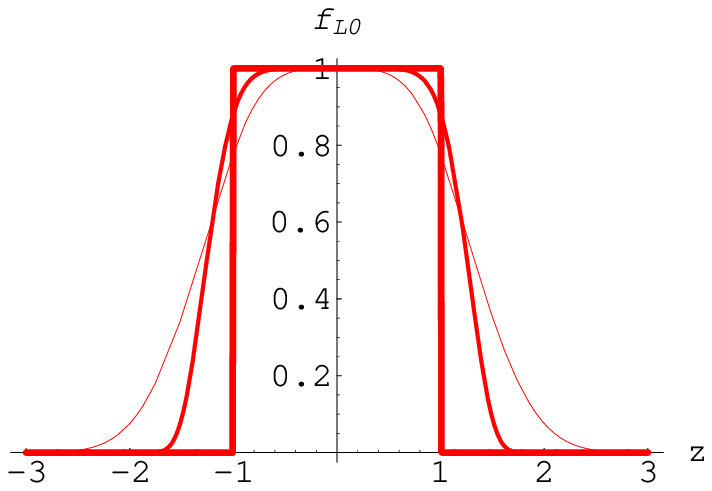}}
\subfigure[$a=0.5,s=1$]{\label{fig_fAL_CaseIIIc}
\includegraphics[width=6cm,height=4cm]{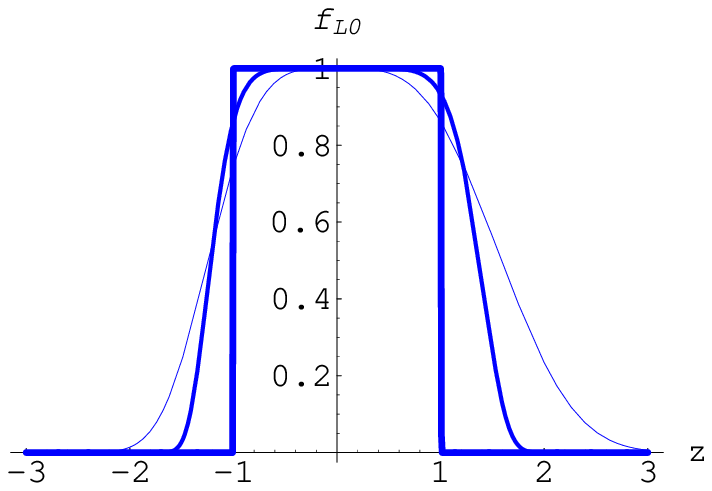}}
\subfigure[$a=0.5,s\rightarrow\infty$]{\label{fig_fAL_CaseIIId}
\includegraphics[width=6cm,height=4cm]{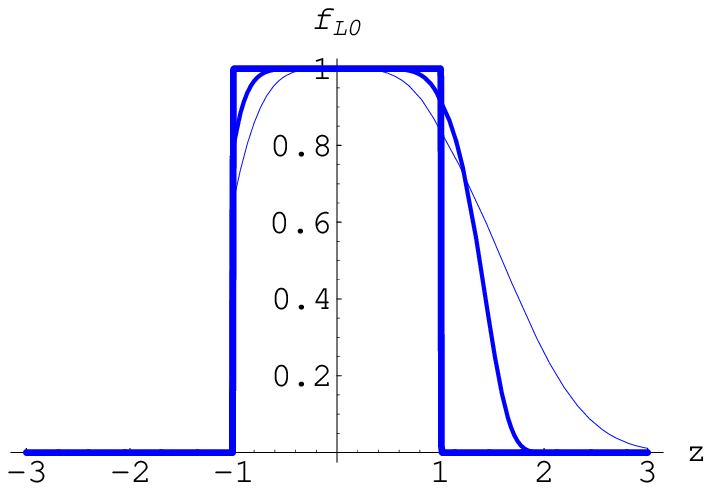}}
\end{center}
\vskip -7mm \caption{(Color online) Comparison of the zero modes of
the asymmetric potential $V^A_L(z)$ (down two) with that of the
symmetric potential $V^S_L(z)$ (up two) for the case
$F(\phi)=\tan^{k/s}(\phi/\phi_0)$ with different $k$ and $s$. The
parameters are set to $\eta=\lambda=1$, $k=3$ for thin lines, $k=7$
for thicker lines, and $k\rightarrow\infty$ for thickest lines.}
 \label{fig_fAL_CaseIII}
\end{figure*}

The normalizable zero mode of left chiral fermions
\begin{equation}
 f_{L0}(z)    \propto
 \exp\left(\int^z d\bar{z}
   \frac{-\eta(\lambda \bar{z})^k
           [1+({\lambda}\bar{z})^{2s}]^{-\frac{1}{2s}}}
        {1+a \bar{z} {~_2F_1}\left(\frac{1}{2 s},\frac{1}{s},
           1+\frac{1}{2s},-({\lambda}\bar{z})^{2s}\right)}
 \right)   \label{zeroModefAL0CaseIII}
\end{equation}
can also be localized on the brane without additional conditions.
The effect of $\eta$, $k$, $s$ and the asymmetric factor $a$ to
the zero mode is shown in Figs. \ref{fig_fAL_CaseIII_eta} and
\ref{fig_fAL_CaseIII}. It can be seen that, with increase of
$\eta$ or $k$, the difference of zero modes with different $a$
would reduce, i.e., the effect of the asymmetric factor could be
neglected. While, with increase of $a$, the effect of $s$ can not
be neglected. For the case $a=0.5$ and $s\rightarrow\infty$, our
four-dimensional massless left fermions can not appear in the
range $z<-1/\lambda$ (see Fig. \ref{fig_fAL_CaseIIId}), namely,
the left sub-brane is the left boundary of the region that the
four-dimensional massless left fermions could appear. The effect
of $k$ to the zero mode is remarkable: with the increase of $k$
(i.e., the increase of the kink-fermion interaction), the region
that the four-dimensional massless left fermions can appear would
reduce. Especially, for the limit case $k\rightarrow\infty$, we
have
\begin{equation}
 f_{L0}(z)    \propto \left\{
 \begin{array}{ll}
   1,  & ~|\lambda {z}|<1 \\
   0, & ~|\lambda {z}|>1
 \end{array}\right. ,   \label{zeroModefAL0CaseIII}
\end{equation}
which shows that the four-dimensional massless left fermions can
only exist in between the locations of two sub-branes and the
probability they would appear is equal everywhere within the
region.

\section{Discussions and conclusions}\label{secConclusion}

In this paper, we have investigated the localization problem and the
spectrum of spin half fermions on a two-parameter ($s$ and
$\lambda$) family of symmetric branes and on a three-parameter ($s$,
$\lambda$ and $a$) family of asymmetric thick branes in an AdS
background for three kinds of typical kink-fermion couplings. The
parameter $a$, which is called the asymmetry factor in this paper,
decides the asymmetry of the solution. The parameter $\lambda$
labels the thickness of the brane. For $s=1$, the solution denotes a
usual thick one-brane, which is the regularized version of the
Randall-Sundrum thin brane. For $a=0$ ($a\neq0$) and odd $s>1$, the
solution stands for a symmetric (an asymmetric) static double brane
interpolating between same (different) AdS$_{5}$ vacua. The
thickness of the double brane is $2/\lambda$, the two sub-branes are
localized at $z=\pm 1/\lambda$. The thickness of the sub-brane
decreases with the increase of $s$. For the limit case
$s\rightarrow\infty$, each thick sub-brane becomes a thin brane.

By presenting the mass-independent potentials (\ref{Vfermion}) of
KK modes in the corresponding Schr\"{o}dinger equations, we
investigate the localization and mass spectra of bulk fermions on
the symmetric and asymmetric thick branes. The formation of the
potentials (\ref{Vfermion}) have two sources: the gravity-fermion
coupling $\bar{\Psi} \Gamma^M \omega_M \Psi$ and the
scalar-fermion coupling $-\eta \bar{\Psi} F(\phi) \Psi$. It can be
seen that, without the gravity-fermion coupling, namely, only
considering domain walls in a flat space-time, the potentials do
not disappear and hence fermions could be localized on the domain
walls (see e.g. \cite{Rubakov1983}). In fact, for a kink solution,
it is clear that the potential for one of left and right chiral
fermions would be a PT like potential, for which the massless mode
of left or right chiral fermion can be localized on the domain
wall without additional condition. However, without scalar-fermion
coupling ($\eta=0$), there is no bound KK mode for both left and
right chiral fermions. Hence, in order to localize the massless
and massive left or right chiral fermions on the branes, some kind
of Yukawa coupling should be introduced.

The spectra are determined by the behavior of the potentials at
infinity. The potentials we are interesting in have three types:\\
 \indent 1) $V_{L,R}(|z|\rightarrow \infty)\rightarrow 0$,\\
 \indent 2) $V_{L,R}(|z|\rightarrow \infty)\rightarrow C$,\\
 \indent 3) $V_{L,R}(|z|\rightarrow \infty)\rightarrow \infty$,\\
where $C$ is a positive constant. In order to realize the three type
of potentials, we have considered three typical Yukawa couplings
correspondingly in this paper:\\
 \indent Case I: `weak' interaction with $F(\phi)=\phi^k$ ($k\geq1$),\\
 \indent Case II: `critical' interaction with
         $F(\phi)=\tan^{1/s}(\phi/\phi_0)$, \\
 \indent Case III: `strong' interaction with
         $F(\phi)=\tan^{k/s}(\phi/\phi_0)$ ($k>1$).\\
Note that, as discussed above, for a domain wall solution in a flat
space-time, a weak kink-fermion interaction would become a strong
interaction. This means that the interaction with gravity would
destroy the localization of fermions on the brane, in a way. So, the
localization of fermions on the brane is the result of the
competition of two interactions.

For the simplest Yukawa coupling with $F(\phi)=\phi$ and $\eta>0$,
the potentials for left chiral fermions provide no mass gap to
separate the fermion zero mode from the excited KK modes. Provided
the condition (\ref{conditionCaseI}), the zero mode of left chiral
fermions can be localized on the brane. The massless mode of left
chiral fermion is most easy to be localized on the symmetric
single brane (i.e., the $a=0$ case). The asymmetric factor $a$ may
destroy the localization of massless fermions. For $s>1$ (the
double brane case), the potential for left chiral fermions have a
double well at the location of the branes. The corresponding zero
mode on both the symmetric and asymmetric double walls is
essentially constant between the two interfaces. This is very
different from the case of gravitons, scalars and vectors, where
the massless modes on the asymmetric double wall are strongly
localized only on the interface centered around the lower minimum
of the potential. The massive KK modes asymptotically turn into
continuous plane waves when far away from the brane, which
represent delocalized massive KK fermions. The massive modes with
lower energy would experience an attenuation due to the presence
of the potential barriers near the location of the brane. It is
interesting to notice that, for $s\geq 3$, a potential well around
the brane location for right chiral fermions would appear and the
well becomes deeper and deeper with increase of $\eta$. We have
shown that this potential would result in a series of massive
fermions with a finite lifetime \cite{0901.3543,Liu0904.1785}. The
spectra of left-handed and right-handed fermionic resonances are
the same, which demonstrates that a Dirac fermion could be
composed from the left and right resonance KK modes
\cite{Liu0904.1785}.

For the critical interaction with
$F(\phi)=\tan^{1/s}(\phi/\phi_0)$ and $\eta>0$, we get a PT-like
potential for left chiral fermions, which provides mass gap to
separate the zero mode from the excited KK modes. The mass spectra
for left and right chiral fermions are almost the same, and the
only one difference is the absent of the zero mode of right chiral
fermions. The massless KK mode of left chiral fermions is
normalizable without additional conditions, and it would be
strongly localized on the brane with large coupling constant
$\eta$. The massive bound KK modes would appear provided large
$\eta$. The spectra for the single brane and the double brane are
quite different. For large $\eta$, there are more bound massive KK
modes on the single brane than on the double brane. For the double
thin brane ($s\rightarrow\infty$), a harmonic oscillator potential
well with finite depth will get for both left and right chiral
fermions and there are finite bound KK modes. For the asymmetric
brane case, the potentials $V^{S}_{L,R}(z)$ are enlarged at
$z\rightarrow +\infty$ and diminished at $z\rightarrow -\infty$,
which shows that the asymmetric factor would reduce the number of
the bound KK modes of left and right fermions. The continuous
spectrum starts at different values for different $s$. The
spectrum structure for the single brane case ($s=1$) is
dramatically changed by the asymmetric factor: the number of bound
KK modes quickly decreases with the increase of $a$ and the
difference with the double brane case is reduced.

For the strong interaction with $F(\phi)=\tan^{k/s}(\phi/\phi_0)$
($k>1$), the potentials $V^{S,A}_{L,R}(z)$ trend to infinite when
far away from the brane and vanish at $z=0$, and there exist
infinite discrete bound KK modes. The influence of $s$ is not
important for symmetric potentials $V^S_{L,R}(z)$, which indicates
that the spectra on the single brane and the double brane are
almost the same. While the increase of $k$ will dramatically
changes the shape of the potentials. Especially, for
$k\rightarrow\infty$, the potential for right hand fermions is an
infinite deep square well. For a fixed finite $k$, the difference
of the symmetric and asymmetric potentials would become large with
the increase of $s$. For $s\rightarrow\infty$, the difference is
largest. While for a fixed $s$, the difference of the two
potentials would become small with the increase of $k$. The
normalizable zero mode of left chiral fermions can also be
localized on the brane without additional conditions. With the
increase of $\eta$ or $k$, the effect of the asymmetric factor to
the zero mode can be neglected. While, with the increase of $a$,
the effect of $s$ is obvious. For the limit case
$s\rightarrow\infty$, the left sub-brane is the left boundary of
the region that the four-dimensional massless left fermions could
appear. With the increase of $k$, the region that the
four-dimensional massless left fermions can appear would reduces.
Especially, for the limit case $k\rightarrow\infty$, the
four-dimensional massless left fermions can only exist in between
the locations of two sub-branes with equal probability.

\section{Acknowledgement}

This work was supported by the Program for New Century Excellent
Talents in University, the National Natural Science Foundation of
China (No. 10705013), the Doctoral Program Foundation of
Institutions of Higher Education of China (No. 20070730055), the
Key Project of Chinese Ministry of Education (No. 109153).

\end{document}